\documentclass[twocolumn]{aastex631}

\defcitealias{XRISM_Perseus}{Velocity~Paper}

\begin{document}

\title{Probable Detection of a Cooler Gas Component in the Perseus Cluster with XRISM}

\author{Julian Meunier}
% \correspondingauthor{Julian Meunier}
\affiliation{Waterloo Centre for Astrophysics, Department of Physics \& Astronomy, University of Waterloo, Ontario N2L 3G1, Canada}
% \email{jjjmeuni@uwaterloo.ca}

% Main Author 2
\author{Brian R. McNamara}
% \correspondingauthor{Brian R. McNamara}
\affiliation{Waterloo Centre for Astrophysics, Department of Physics \& Astronomy, University of Waterloo, Ontario N2L 3G1, Canada}
% \email{mcnamara@uwaterloo.ca}

% Discussion, modeling, systematics (thank you!)
\author{Aurora Simionescu}
\affiliation{SRON Space Research Organisation Netherlands, Niels Bohrweg 4, 2333 CA, Leiden, The Netherlands}
% \email{a.simionescu@sron.nl}

% Discussion, modeling, abundances (thank you!)
\author{Fran\c cois Mernier}
\affiliation{Department of Astronomy, University of Maryland, College Park, MD 20742, USA}
\affiliation{NASA / Goddard Space Flight Center, Greenbelt, MD 20771, USA}
\affiliation{Center for Research and Exploration in Space Science and Technology, NASA / GSFC (CRESST II), Greenbelt, MD 20771, USA}
% \email{francois.mernier@irap.omp.eu}

% Responsible for the bin1 data/rmfs/arfs used, and object PI. Draft comments & discussion (thank you!)
\author{Irina Zhuravleva}
\affiliation{Department of Astronomy and Astrophysics, University of Chicago, Chicago, IL 60637, USA}
% \email{zhuravleva@uchicago.edu}

% Responsible for Line_Remover model. Draft comments & discussion (thank you!)
\author{Congyao Zhang}
\affiliation{Department of Theoretical Physics and Astrophysics, Masaryk University, Brno 61137, Czechia}
\affiliation{Department of Astronomy and Astrophysics, University of Chicago, Chicago, IL 60637, USA}
% \email{cyzhang@astro.uchicago.edu}

% Draft comments & discussion (thank you!)
\author{Annie Heinrich}
\affiliation{Department of Astronomy and Astrophysics, University of Chicago, Chicago, IL 60637, USA}
% \email{amheinrich@uchicago.edu}

% Draft comments (thank you!)
\author{Julie Hlavacek-Larrondo}
\affiliation{D\'epartement de Physique, Universit\'e de Montr\'eal, Succ. Centre-Ville, Montr\'eal, Qu\'ebec H3C 3J7, Canada}
% \email{j.larrondo@umontreal.ca}

% Draft comments (thank you!)
\author{Frederick S. Porter}
\affiliation{NASA / Goddard Space Flight Center, Greenbelt, MD 20771, USA}
% \email{Frederick.S.Porter@nasa.gov}

% Draft comments (thank you!)
\author{Benjamin Vigneron}
\affiliation{D\'epartement de Physique, Universit\'e de Montr\'eal, Succ. Centre-Ville, Montr\'eal, Qu\'ebec H3C 3J7, Canada}
% \email{benjamin.vigneron@umontreal.ca}

% Draft comments (thank you!)
\author{John ZuHone}
\affiliation{Center for Astrophysics $\vert$ Harvard \& Smithsonian, Cambridge, MA 02138, USA}
% \email{john.zuhone@cfa.harvard.edu}

\author{Elena Bellomi}
\affiliation{Center for Astrophysics $\vert$ Harvard \& Smithsonian, Cambridge, MA 02138, USA}
% \email{elena.bellomi@cfa.harvard.edu}

\author{Ian Drury}
\affiliation{Department of Physics and Astronomy, The University of Georgia, Athens, GA 30602, USA}
% \email{ian.drury@uga.edu}

\author{Megan E. Eckart}
\affiliation{Lawrence Livermore National Laboratory, CA 94550, USA}
% \email{eckart2@llnl.gov}

\author{Ryuichi Fujimoto}
\affiliation{Institute of Space and Astronautical Science (ISAS), Japan Aerospace Exploration Agency (JAXA), Kanagawa 252-5210, Japan}
% \email{fujimoto.ryuichi@jaxa.jp}

\author{Yutaka Fujita}
\affiliation{Department of Physics, Tokyo Metropolitan University, Tokyo 192-0397, Japan}
% \email{y-fujita@tmu.ac.jp}

\author{Liyi Gu}
\affiliation{SRON Netherlands Institute for Space Research, Leiden, The Netherlands}
% \email{L.Gu@sron.nl}

\author{Isamu Hatsukade}
\affiliation{Faculty of Engineering, University of Miyazaki, Miyazaki 889-2192, Japan}
% \email{hatukade@cs.miyazaki-u.ac.jp}

\author{Yuto Ichinohe}
\affiliation{RIKEN Nishina Center, Saitama 351-0198, Japan}
% \email{ichinohe@ribf.riken.jp}

\author{Yoshiaki Kanemaru}
\affiliation{Institute of Space and Astronautical Science (ISAS), Japan Aerospace Exploration Agency (JAXA), Kanagawa 252-5210, Japan}
% \email{kanemaru.yoshiaki@jaxa.jp}

\author{Takao Kitaguchi}
\affiliation{RIKEN Nishina Center, Saitama 351-0198, Japan}
% \email{takao.kitaguchi@riken.jp}

\author{Shunji Kitamoto}
\affiliation{Department of Physics, Rikkyo University, Tokyo 171-8501, Japan}
% \email{skitamoto@rikkyo.ac.jp}

\author{Shogo Kobayashi}
\affiliation{Faculty of Physics, Tokyo University of Science, Tokyo 162-8601, Japan}
% \email{shogo.kobayashi@rs.tus.ac.jp}

\author{Takayoshi Kohmura}
\affiliation{Faculty of Science and Technology, Tokyo University of Science, Chiba 278-8510, Japan}
% \email{tkohmura@rs.tus.ac.jp}

\author{Hironori Matsumoto}
\affiliation{Department of Earth and Space Science, Osaka University, Osaka 560-0043, Japan}
% \email{matumoto@ess.sci.osaka-u.ac.jp}

\author{Kyoko Matsushita}
\affiliation{Faculty of Physics, Tokyo University of Science, Tokyo 162-8601, Japan}
% \email{matusita@rs.tus.ac.jp}

\author{Kostas Migkas}
\affiliation{SRON Netherlands Institute for Space Research, Leiden, The Netherlands}
% \email{k.migkas@sron.nl}

\author{Ikuyuki Mitsuishi}
\affiliation{Department of Physics, Nagoya University, Aichi 464-8602, Japan}
% \email{ikuyuki.mitsuishi@gmail.com}

\author{Koji Mori}
\affiliation{Faculty of Engineering, University of Miyazaki, Miyazaki 889-2192, Japan}
% \email{mori@astro.miyazaki-u.ac.jp}

\author{Hiroshi Nakajima}
\affiliation{College of Science and Engineering, Kanto Gakuin University, Kanagawa 236-8501, Japan}
% \email{hiroshi@kanto-gakuin.ac.jp}

\author{Hirofumi Noda}
\affiliation{Astronomical Institute, Tohoku University, Miyagi 980-8578, Japan}
% \email{hirofumi.noda@astr.tohoku.ac.jp}

\author{Anna Ogorzalek}
\affiliation{Department of Astronomy, University of Maryland, College Park, MD 20742, USA}
\affiliation{NASA / Goddard Space Flight Center, Greenbelt, MD 20771, USA}
\affiliation{Center for Research and Exploration in Space Science and Technology, NASA / GSFC (CRESST II), Greenbelt, MD 20771, USA}
% \email{ogoann@umd.edu}

\author{Naomi Ota}
\affiliation{Department of Physics, Nara Women’s University, Nara 630-8506, Japan}
% \email{naomi@cc.nara-wu.ac.jp}

\author{Lior Shefler}
\affiliation{Department of Physics and Astronomy, The University of Georgia, Athens, GA 30602, USA}
% \email{liorshefler@uga.edu}

\author{Nhut Truong}
\affiliation{Academia Sinica Institute of Astronomy and Astrophysics (ASIAA), Taipei, 106319, Taiwan}
\affiliation{Center for Astrophysics $\vert$ Harvard \& Smithsonian, Cambridge, MA 02138, USA}
\affiliation{Center for Research and Exploration in Space Science and Technology, NASA / GSFC (CRESST II), Greenbelt, MD 20771, USA}
% \email{ntruong@umbc.edu}

\author{Ayşegül Tümer}
\affiliation{Center for Space Science and Technology, University of Maryland, Baltimore County (UMBC), Baltimore, MD 21250, USA}
\affiliation{NASA / Goddard Space Flight Center, Greenbelt, MD 20771, USA}
\affiliation{Center for Research and Exploration in Space Science and Technology, NASA / GSFC (CRESST II), Greenbelt, MD 20771, USA}
% \email{aysegultumer@gmail.com}

\author{Nagomi Uchida}
\affiliation{Institute of Space and Astronautical Science (ISAS), Japan Aerospace Exploration Agency (JAXA), Kanagawa 252-5210, Japan}
% \email{uchida.nagomi@jaxa.jp}

\author{Yuusuke Uchida}
\affiliation{Faculty of Science and Technology, Tokyo University of Science, Chiba 278-8510, Japan}
% \email{uchida.yuusuke99@jaxa.jp}

\author{Shutaro Ueda}
\affiliation{Faculty of Mathematics and Physics, Institute of Science and Engineering, Kanazawa University, Kakuma, Kanazawa, Ishikawa, 920-1192, Japan}
\affiliation{Advanced Research Center for Space Science and Technology, Kanazawa University, Kakuma, Kanazawa, Ishikawa, 920-1192, Japan}
\affiliation{Academia Sinica Institute of Astronomy and Astrophysics (ASIAA), Taipei, 106319, Taiwan}
% \email{shutaro@se.kanazawa-u.ac.jp}



\begin{abstract}

We present an analysis of the temperature structure of the Perseus cluster atmosphere using XRISM Resolve observations. The average temperature rises from 3.3 keV near the nucleus of NGC 1275 to 8 keV at 10 arcmin (210 kpc), which is consistent with Chandra and XMM measurements.  The velocity and velocity dispersion profiles are broadly consistent with those in \citet{XRISM_Perseus}.  While the gas at altitudes beyond $\sim60$ kpc can be modeled as a single temperature plasma, we find evidence for more than one gas phase in the inner $\sim60$ kpc. The hotter gas component, traced primarily by the Fe He$\alpha$ line, has a velocity dispersion of $\lesssim140$ km s$^{-1}$. We detect a cooler, $\sim1.87-2.43$ keV, gas component with a velocity dispersion of $\sim300-400$ km s$^{-1}$ and a bulk velocity of $\sim 21-213$ km s$^{-1}$ with respect to the central galaxy. These ranges reflect large systematic uncertainties associated with modeling spatial-spectral mixing and the bright central point source. Potential low energy gain variations may add additional uncertainties. The cooler component is identified by broad wings in prominent emission lines, most notably S Ly$\alpha$ and Fe He$\alpha$. This cooler component's Mach number $\sim0.73-0.96$ and non-thermal pressure fraction of $\sim22.9-33.7\%$ are much higher than found for the hotter gas. The cooler gas may be associated with merging halos along the line of sight which formed the cool, sloshing spiral and/or cooling gas being disturbed by the radio jets and lobes.

\end{abstract}



\section{Introduction}\label{Intro}
Galaxy clusters are surrounded by massive atmospheres of low density hot gas, typically very bright in X-ray (e.g. \citealt{Allen2001,McNamaraNulsen2007,Fabian2011}). The feedback between the intracluster medium (ICM) and the active galactic nuclei (AGN) of the brightest central galaxy (BCG) is a crucial physical process to the evolution of these clusters and surrounding baryonic matter (e.g. \citealt{Voit2005,McNamaraNulsen2012}). The Perseus cluster is the brightest nearby X-ray galaxy cluster, featuring cavities, shocks, ripples, and sloshing gas \citep{Fabian2003,FabianPerseus2006,FabianPerseus2007}. The supermassive black hole in the BCG, NGC 1275, interacts with the ICM by inflating bubbles of relativistic plasma, seen in the form of X-ray cavities. These bubbles rise through the ICM due to buoyancy, displacing the hot gas and perhaps sowing turbulence \citep{Zhuravleva2014}. Evidence of ongoing mergers from infalling galaxies is seen, causing sloshing motions of the gas seen in the form of spiral structures in the X-ray residual image (e.g. \citealt{Churazov2003,Zuhone2011,Simionescu2012}).

Perseus has previously been observed with Hitomi \citep{Perseus_Hitomi_2016}, detecting velocity dispersions of $<200$ km s$^{-1}$ in the inner $\sim60$ kpc. Further efforts were made to extract multiple temperature structure in the inner regions using emission line ratios and multiple temperature modeling \citep{Hitomi_Perseus_lineratios}. A combination of 3 keV and 5 keV gas was detected providing potential evidence for multiple temperature structure in the core region of the cluster, but attributed to projection of the radial temperature gradient along the line of sight. 

Perseus has been extensively observed with Chandra, which confirmed the cavities and identified shock fronts and sloshing gas in the cluster core \citep{Fabian2003,FabianPerseus2006,FabianPerseus2007}. The gas in the inner $\sim60$ kpc has multiple phases, having different distributions of mass at different temperatures (0.5--4 keV) in the plane of the sky, including a sloshing gas spiral structure at 2 keV, demonstrating that the multiple temperature nature of the gas is not just a projection effect \citep{FabianPerseus2006}.

\begin{figure*}[ht]
    \centering
    \includegraphics[width=0.75\textwidth, keepaspectratio]{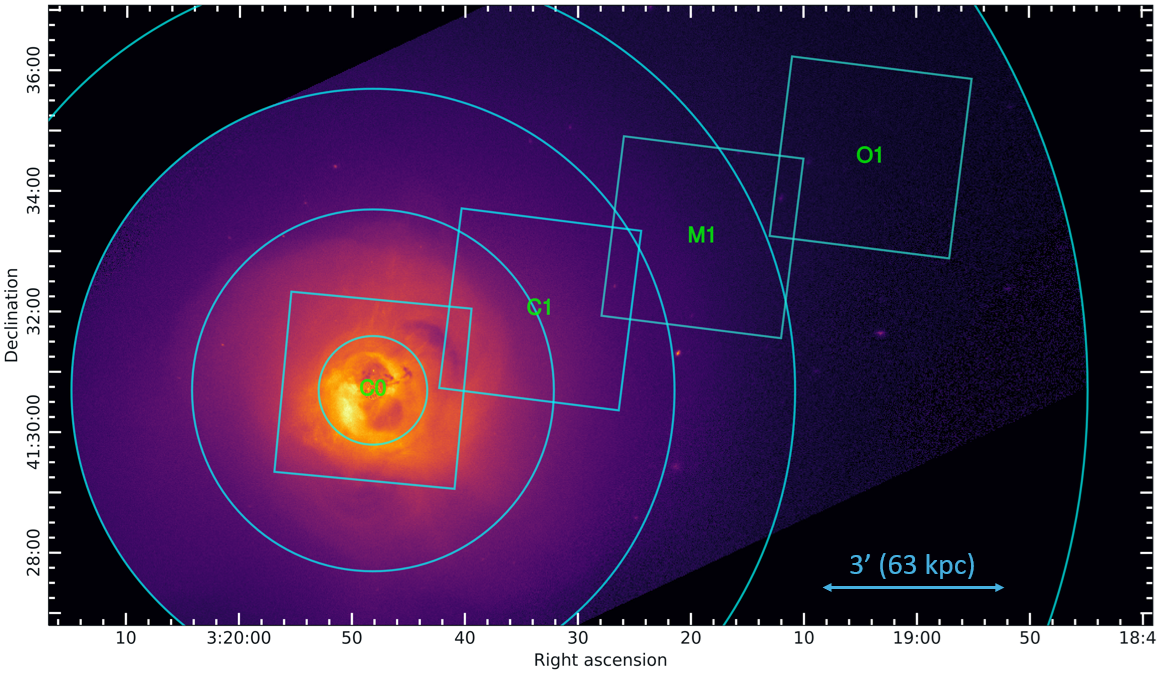}
    \includegraphics[width=0.75\textwidth, keepaspectratio]{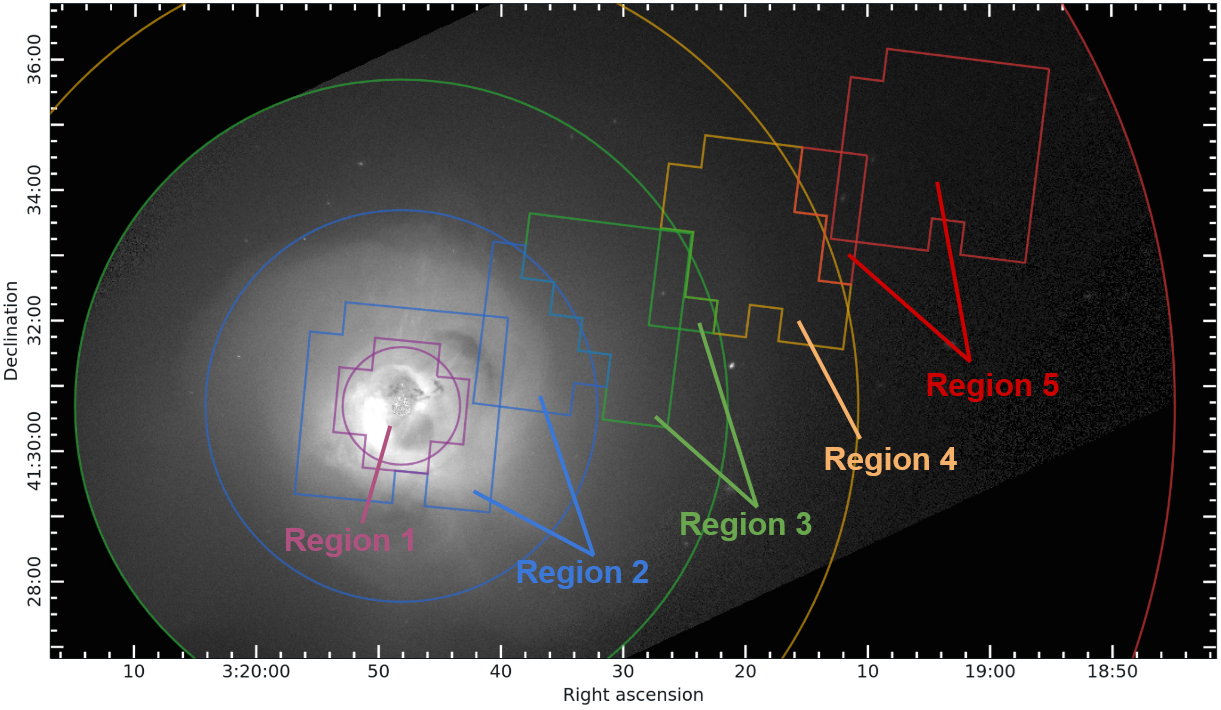}
    \caption{\textit{Top:} The four XRISM pointings (squares) and the five mapped regions (circles) of the Perseus cluster overlaid on the cropped exposure-corrected 2--8 keV Chandra ACIS image \citep{XRISM_Perseus}. \textit{Bottom:} The five pixel sub-FOV regions of the multiple temperature fitting model.}
    \label{fig:pointings}
\end{figure*}

The X-Ray Imaging and Spectroscopy Mission (XRISM) launched in September 2023, with a high resolution microcalorimeter, Resolve, on board. XRISM has already observed multiple clusters, including Abell 2029 \citep{XRISM_2025_A2029}, Centaurus \citep{XRISM_Centaurus_2025}, Coma \citep{XRISM2025Coma}, Cygnus A \citep{XRISM_CygnusA}, Hydra A \citep{XRISM2025HydraA}, and Ophiuchus \citep{XRISM_Ophiuchus2025}. These clusters report relatively low central velocity dispersions of $\sim260$ km s$^{-1}$ or less. Abell 2029 reported a two-temperature model to account for the cool core. However, the velocity dispersion ($\sim165$ km s$^{-1}$) was tied between the two components, and both components are relatively hot ($\gtrsim 5$ keV) \citep{XRISM_2025_A2029}. Cygnus A shows evidence for a cooler gas component ($\sim2$ keV) with a broad velocity dispersion of $\sim440\pm130$ km s$^{-1}$, interpreting the hotter and cooler components as turbulence and motion of the cocoon shock \citep{XRISM_CygnusA}.

XRISM has also observed the Perseus cluster (\citealt{XRISM_Perseus}, henceforth \citetalias{XRISM_Perseus}), measuring radial profiles of temperature and velocity dispersions up to $\sim250$ kpc along the NW direction from the cluster center. Low velocity dispersion of $\simeq 135-172$ km s$^{-1}$ is measured in the inner $\sim50$ kpc. A radially-declining velocity dispersion gradient in the inner $\sim60$ kpc suggests the jets and bubbles are driving turbulence in the core \citepalias{XRISM_Perseus}. However, the level of heating in the inner 100 kpc tends to lie below the cooling rate for reasonable assumptions about the injection scales indicating that turbulence would struggle to balance cooling (cf., \citealt{XRISM2025HydraA}). 

The temperature of the gas can be estimated by the flux ratios of different emission lines. Flux ratios of different transitions in the same ion can be used to estimate the excitation and ionization temperatures of that species in the plasma. Flux ratios of different ionization species represent the ion fraction of each element \citep{Hitomi_Perseus_lineratios}. These measurements should match the gas temperature derived from the continuum shape if the plasma is a single temperature gas in collisional ionization equilibrium (CIE). A difference between the temperature estimated using flux ratios and the continuum shape may indicate a complex temperature structure or that the plasma is out of collisional ionization equilibrium. Deviations due to multiple temperatures can be modeled. The high spectral resolution of Resolve ($\Delta\mathrm{E}\sim5$ eV at 6.4 keV) allows for the precise measurement of fluxes of the brightest lines and potential multiple temperature structure from features of the emission lines or the continuum shape.

Here we examine the temperature structure of the atmosphere through emission line ratios and multiple temperature spectral modeling. In this paper, we assume a Hubble constant of H$_{0}= 70$ km s$^{-1}$ Mpc$^{-1}$. We assume the cluster redshift to be $z = 0.017284$ \citep{Perseus_Hitomi_2018}, which gives a spatial scale of 21 kpc/arcmin. Reported errors are statistical $1\sigma$ errors unless stated otherwise.

\section{Observations and Data Reduction}\label{Methods}
The Perseus cluster was observed five times with XRISM between 21 January and 26 January, 2024. Two observations were pointed towards the cluster center (Obs IDs: 000154000, 000155000), and three were pointed along the north-west direction (Obs IDs: 000156000, 000157000, 000158000). Figure \ref{fig:pointings} shows the footprint labeled C0, C1, M1, and O1 from the cluster center outward. The Resolve data was reprocessed with the XRISM pre-release software build 8, and calibrated using CalDB version 8. The data was screened according to the XRISM Quick Start Guide Version 2.1, keeping only the highest resolution primary events. The resulting exposure times for Obs IDs 000154000 -- 000158000 after reprocessing are 48, 50, 57, 93, and 131 ks. Pixel 27 was excluded due to unexpected gain drifts \citep{XRISM_flightdata2025}, and pixel 12 was excluded as it is the calibration pixel. The spectral resolution of the Resolve data after reprocessing is $\simeq4.5$ eV FWHM.

Redistribution matrix files (RMFs) and auxiliary response files (ARFs) were generated with XRISM pre-release software build 8. Extra-large (`X') size RMFs were generated with \texttt{rslmkrmf} which were used in this analysis, which are necessary for spectral fitting below $\sim2.4$ keV. ARFs were generated with \texttt{xaexpmap} and \texttt{xaarfgen}. The 2--8 keV exposure-corrected Chandra ACIS image (Fig. \ref{fig:pointings}) described in the \citetalias{XRISM_Perseus} was used as the source model for ARF generation for all pointings and sub-FOV regions. As the image has the AGN masked out, a point source model was used to generate ARFs to account for contributions from the bright AGN source.

For the multiple temperature analysis, the pointings were divided into 5 sub-FOV regions, shown in Figure \ref{fig:pointings}, for finer spatial detail. A spectrum was processed for each relevant sub-FOV region for each pointing, for a total of 10 spectra. RMFs and ARFs were generated for each region. Additional ARFs were generated for each region accounting for photon contamination from other regions, as described in the \citetalias{XRISM_Perseus}, which we will refer to as spatial-spectral mixing (SSM).

\section{Analysis}\label{Analysis}
Spectral fits were done with Xspec 12.14.1 \citep{Xspec}, using the Cash statistic \citep{Cash} to statistically evaluate the quality of the fit. We used the atomic database \texttt{AtomDB} v3.0.9 \citep{atomdb} for calculating the plasma models. The plasma model we used is the collisional ionization equilibrium model, \texttt{apec}, and its variants for temperature distribution, velocity broadening, metallicities, and multiple temperatures. We used the Tuebingen-Boulder absorption model \citep{tbabs}, \texttt{TBabs}, to model the X-ray absorption of the ISM. The hydrogen column density $\text{N}_{\text{H}} = 1.38\times10^{21}$ cm$^{-2}$ was fixed. All abundances are relative to proto-solar values measured in \cite{Lodders2009}. The plasma model abundance parameters that are not free during fitting are frozen at solar values.

Non-thermal contributions to the spectra from the AGN were modeled with a power law. When left free to vary, the power law favors high photon index, $\Gamma$, and normalization (or flux) in tension with observations, which can cause the ICM models to become unrealistic. Stricter constraints are imposed on the power law for each analysis to properly model the AGN and ICM. For the emission line ratios analysis, the photon index is frozen to the best fit value from the \citetalias{XRISM_Perseus} and the normalization is frozen to the broadband best fit value of the respective pointing. For the multiple temperature analysis, the photon index is free to vary and the flux is frozen to the best fit value from the \citetalias{XRISM_Perseus}. Specific details for each analysis are provided in sections \ref{subsec:Line Ratios} and \ref{subsec:multi-T}. Systematic effects due to the AGN are discussed in section \ref{subsec:systematics:agn}.

Contributions from the non-X-ray background (NXB) were included in all analyses using a diagonal RMF with no ARF. NXB spectra were extracted for each pointing from Resolve's night-Earth database with \texttt{rslnxbgen}, which were modeled empirically. The empirical model consists of a continuum modeled by a power law, and 13 detector background emission lines in the 1.8--10 keV band modeled by gaussians. The best fit models were frozen and included in fitting, applied to the appropriate pointings and sub-FOV regions. For sub-FOV regions, the model was scaled by the size of the extraction region to avoid over-accounting for the NXB. The NXB is typically $\sim10^{-3}$ counts s$^{-1}$ keV$^{-1}$ or less. For the central pointings the NXB only provides significant contributions at the edges of the spectra, but for the outer pointings it is significant throughout the entire band.

\subsection{Emission Line Ratios}\label{subsec:Line Ratios}
\begin{figure}[t]
    \centering
    \includegraphics[width=\columnwidth]{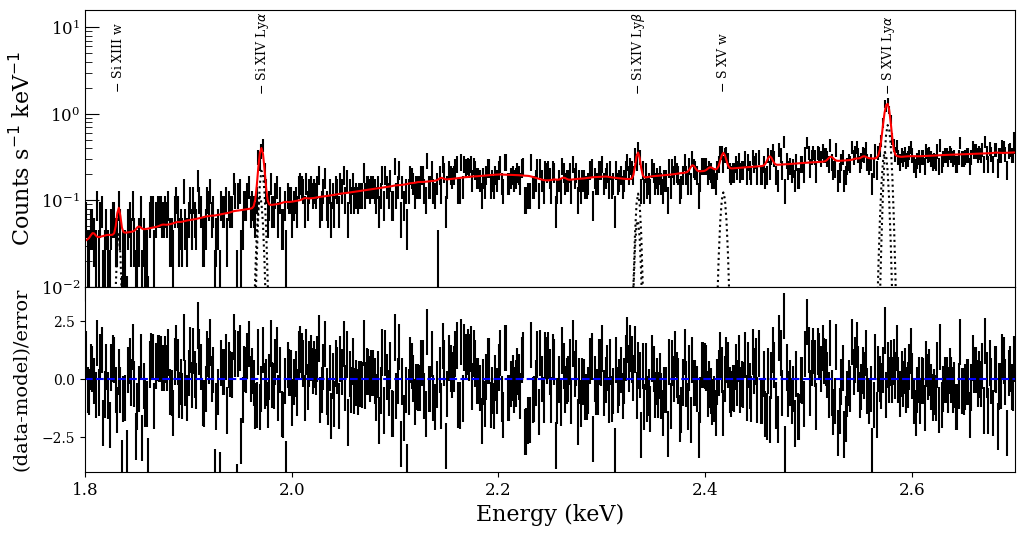}
    \includegraphics[width=\columnwidth]{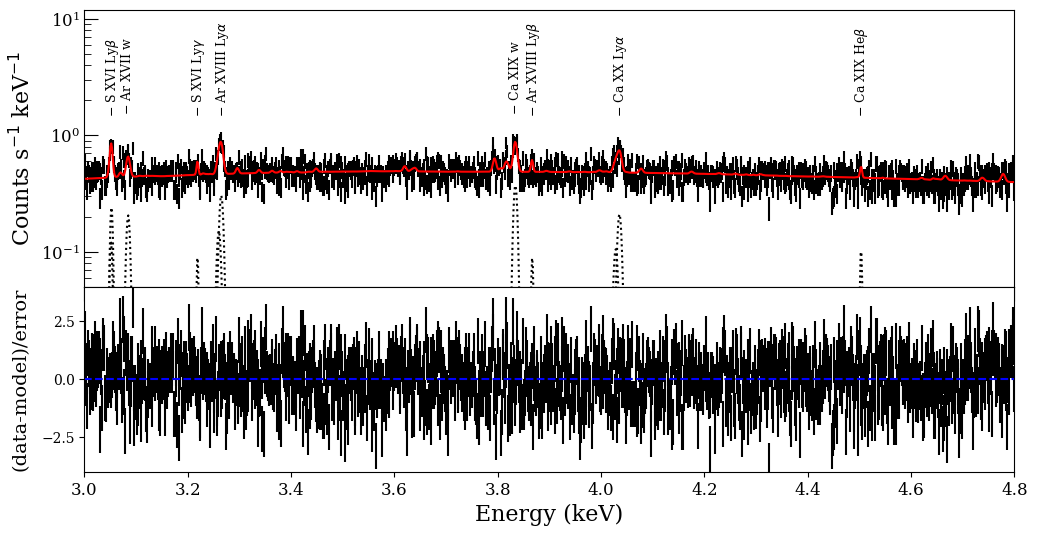}
    \includegraphics[width=\columnwidth]{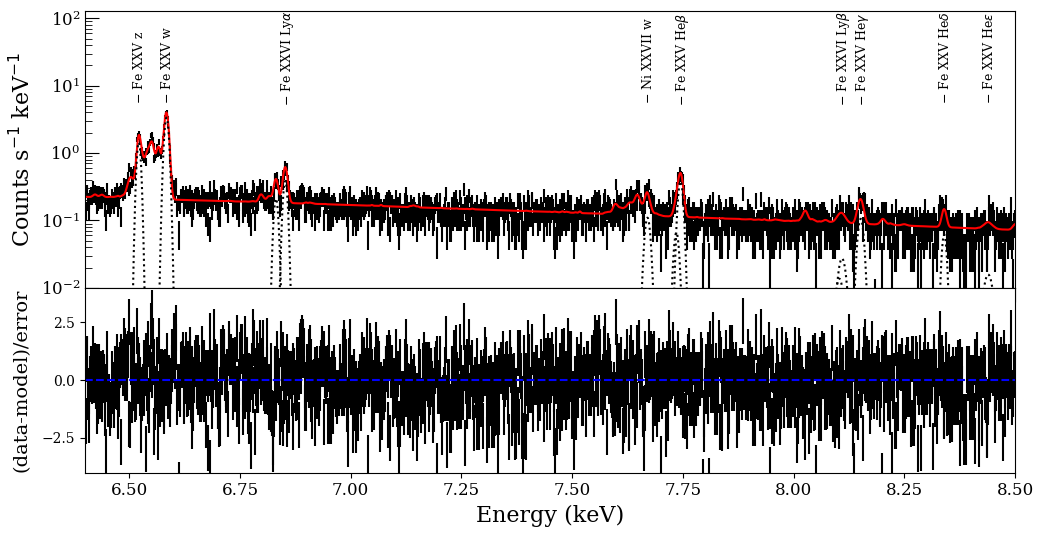}
    \caption{Narrow band fits of the C0 spectrum for 1.8--2.7, 3.0--4.8, and 6.4--8.5 keV. The best fit model is shown in red, and emission lines of interest are labeled. Residuals, calculated as (data - model)/error in terms of $\sigma$, are shown below each fit. The spectrum is binned for plotting purposes only.}
    \label{fig:lineratios_spectrum_plots}
\end{figure}

We performed fits on 1.8--2.7, 3.0--4.8, and 6.4--8.5 keV bands for the C0 and C1 pointings. Figure \ref{fig:lineratios_spectrum_plots} shows the narrow band fits of the C0 spectrum. We detect Ly$\alpha$ \& $\beta$, and He$\alpha$ \& $\beta$ lines for Si, S, Ar, Ca, and Fe. Additionally, we detect Ly$\gamma$ lines for S, and He$\gamma$-$\epsilon$ lines for Fe.

To measure the observed fluxes of the emission lines, we created an alternate version of the \texttt{apec} emission line FITS table in Xspec where the target emission lines were removed. Fluxes are provided in Table \ref{tab:perseus_lines}. The Lyman series doublets, with the exception of Fe {\scriptsize XXVI} Ly$\alpha$, are unresolved, so the gaussian centroids, widths, and relative normalizations are tied to each other (see columns 4--6 of Table \ref{tab:perseus_lines}). Some line fluxes were too low to measure fine structure and were modeled with a single gaussian.

The narrow bands were modeled with \texttt{TBabs*(bapec + (N $\times$ zgauss) + powerlaw) + NXB}, where each narrow band fit has \texttt{N} red-shifted gaussians according to Table \ref{tab:perseus_lines}. The \texttt{bapec} parameters ($k_BT$, abundance, and $\sigma$) and the \texttt{powerlaw} normalization were frozen to the best fit value of a single-temperature fit for each pointing. The \texttt{powerlaw} photon index was fixed at $\Gamma=1.7$. These parameters were fixed to ensure the model does not significantly vary across narrow bands while still allowing freedom for the gaussian components. The frozen parameters are provided in Table \ref{tab:lineratios_base} in appendix \ref{appendix:LineRatios}. The redshifts of the gaussians are all tied to the redshift of the \texttt{bapec} model. The best-fit gaussian parameters of each emission line for both pointings are provided in Table \ref{tab:perseus_lines_results}. Some lines or doublets that were detectable in C0 had insufficient signal in C1 and were either omitted or the doublet replaced with a single gaussian.

\begin{figure}
    \centering
    \includegraphics[width=\linewidth]{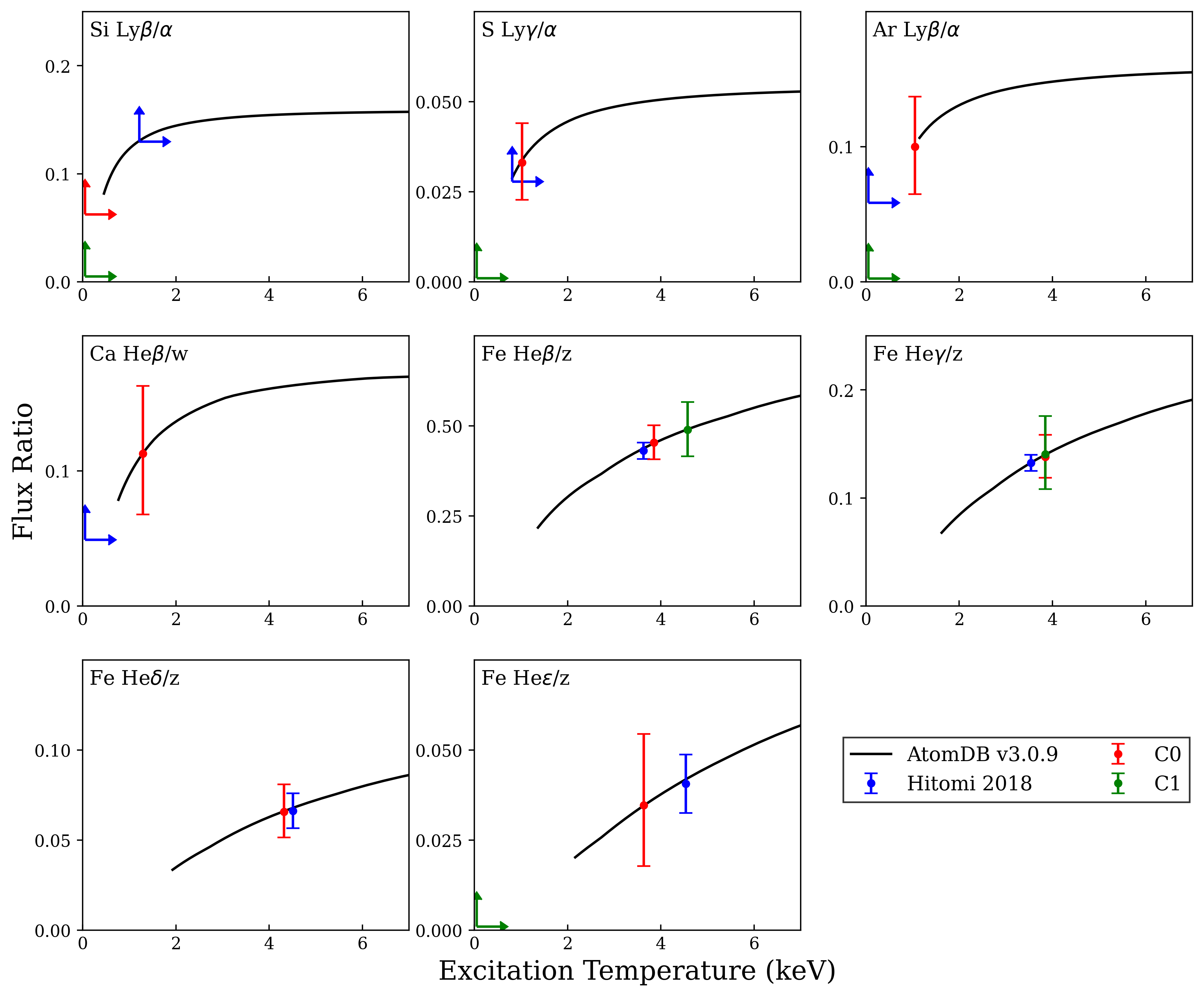}
    \includegraphics[width=\linewidth]{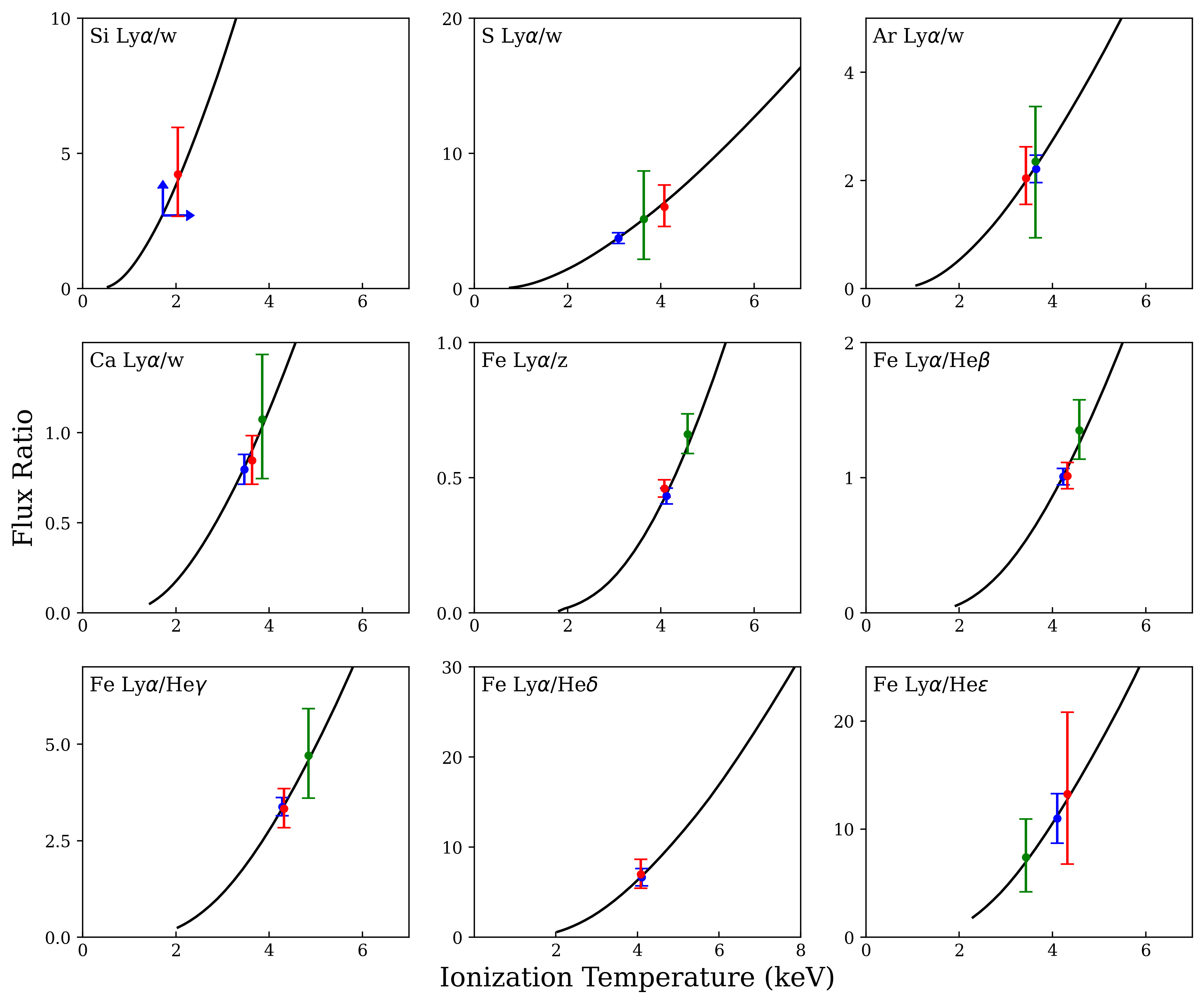}
    \caption{Measured flux ratios for Si, S, Ar, Ca, and Fe lines with respect to theoretical curves from \texttt{AtomDB} v3.0.9. Data points with arrows represent $3\sigma$ lower limits. \textit{Top}: The calculated excitation temperatures from the flux ratios of different transitions of the same ion. \textit{Bottom}: The calculated ionization temperatures from the flux ratios of different ionization species.}
    \label{fig:line_ratios}
\end{figure}

The emission line ratios for Si, S, Ar, Ca, and Fe were calculated for both pointings. The excitation temperature is determined by the flux ratio of various transitions within the same ion and the ionization temperature by the flux ratio of the different ionization species
\citep{Hitomi_Perseus_lineratios}. The fluxes of the Ly doublets are summed as are those with unresolved sub-structure. When a doublet is replaced with a single gaussian in the C1 pointing, the single gaussian flux is used. These ratios are then compared to theoretical curves from \texttt{AtomDB} v3.0.9, shown in Figure \ref{fig:line_ratios}.

\begin{table*}
    \centering
    \caption{Emission lines used in spectral fits, separated by narrow band. Free parameters are indicated by a hyphen (--).}
    \renewcommand{\arraystretch}{1.15}
    \begin{tabular*}{\linewidth}{@{\extracolsep{\fill}} lccccc}
        \hline
         \multicolumn{2}{c}{Emission Line Information} & \multicolumn{4}{c}{Parameter Links} \\
        \hline
        Line name & \multicolumn{1}{c}{$E_0$ (eV)} & Tied to & $E_0$ & $\sigma$ & Norm \\
        \hline
        \hline
        Si {\scriptsize XIII} w & 1865.0 & -- & -- & -- & -- \\
        Si {\scriptsize XIV} Ly$\alpha_1$ & 2006.1 & -- & -- & -- & -- \\
        Si {\scriptsize XIV} Ly$\alpha_2$ & 2004.3 & Si {\scriptsize XIV} Ly$\alpha_1$ & $-1.8$ eV & $\times1.0$ & $\times0.5$ \\
        Si {\scriptsize XIV} Ly$\beta_1$ & 2376.6 & -- & -- & -- & -- \\
        Si {\scriptsize XIV} Ly$\beta_2$ & 2376.1 & Si {\scriptsize XIV} Ly$\beta_1$ & $-0.5$ eV & $\times1.0$ & $\times0.5$ \\
        S {\scriptsize XV} w & 2460.6 & -- & -- & -- & -- \\
        S {\scriptsize XVI} Ly$\alpha_1$ & 2622.7 & -- & -- & -- & -- \\
        S {\scriptsize XVI} Ly$\alpha_2$ & 2619.7 & S {\scriptsize XVI} Ly$\alpha_1$ & $-3.0$ eV & $\times1.0$ & $\times0.5$ \\
        \hline
        S {\scriptsize XVI} Ly$\beta_1$ & 3106.7 & -- & -- & -- & -- \\
        S {\scriptsize XVI} Ly$\beta_2$ & 3105.8 & S {\scriptsize XVI} Ly$\beta_1$ & $-0.9$ eV & $\times1.0$ & $\times0.5$ \\
        S {\scriptsize XVI} Ly$\gamma_1$ & 3276.3 & -- & -- & -- & -- \\
        S {\scriptsize XVI} Ly$\gamma_2$ & 3275.9 & S {\scriptsize XVI} Ly$\gamma_1$ & $-0.4$ eV & $\times1.0$ & $\times0.5$ \\
        Ar {\scriptsize XVII} w & 3139.6 & -- & -- & -- & -- \\
        Ar {\scriptsize XVIII} Ly$\alpha_1$ & 3323.0 & -- & -- & -- & -- \\
        Ar {\scriptsize XVIII} Ly$\alpha_2$ & 3318.2 & Ar {\scriptsize XVIII} Ly$\alpha_1$ & $-4.8$ eV & $\times1.0$ & $\times0.5$ \\
        Ar {\scriptsize XVIII} Ly$\beta_1$ & 3935.7 & -- & -- & -- & -- \\
        Ar {\scriptsize XVIII} Ly$\beta_2$ & 3934.3 & Ar {\scriptsize XVIII} Ly$\beta_1$ & $-1.4$ eV & $\times1.0$ & $\times0.5$ \\
        Ca {\scriptsize XIX} w & 3902.4 & -- & -- & -- & -- \\
        Ca {\scriptsize XIX} He${\beta_1}^\dagger$ & 4583.5 & -- & -- & -- & -- \\
        Ca {\scriptsize XX} Ly$\alpha_1$ & 4107.5 & -- & -- & -- & -- \\
        Ca {\scriptsize XX} Ly$\alpha_2$ & 4100.1 & Ca {\scriptsize XX} Ly$\alpha_1$ & $-7.4$ eV & $\times1.0$ & $\times0.5$ \\
        \hline
        Fe {\scriptsize XXV} z & 6636.6 & -- & -- & -- & -- \\
        Fe {\scriptsize XXV} w & 6700.4 & -- & -- & -- & -- \\
        Fe {\scriptsize XXV} He$\beta_1$ & 7881.5 & -- & -- & -- & -- \\
        Fe {\scriptsize XXV} He$\beta_2$ & 7872.0 & Fe {\scriptsize XXV} He$\beta_1$ & $-9.5$ eV & $\times1.0$ & -- \\
        Fe {\scriptsize XXV} He${\gamma_1}^\dagger$ & 8295.5 & -- & -- & -- & -- \\
        Fe {\scriptsize XXV} He${\delta_1}^\dagger$ & 8487.4 & -- & -- & -- & -- \\
        Fe {\scriptsize XXV} He${\epsilon_1}^\dagger$ & 8588.5 & -- & -- & -- & -- \\
        Fe {\scriptsize XXVI} Ly$\alpha_1$ & 6973.1 & -- & -- & -- & -- \\
        Fe {\scriptsize XXVI} Ly$\alpha_2$ & 6951.9 & Fe {\scriptsize XXVI} Ly$\alpha_1$ & -- & $\times1.0$ & -- \\
        Fe {\scriptsize XXVI} Ly$\beta_1$ & 8252.6 & -- & -- & -- & -- \\
        Fe {\scriptsize XXVI} Ly$\beta_2$ & 8246.4 & Fe {\scriptsize XXVI} Ly$\beta_1$ & $-6.2$ eV & $\times1.0$ & $\times0.5$ \\
        Ni {\scriptsize XXVII} w & 7805.6 & -- & -- & -- & -- \\
        \hline
        \multicolumn{6}{l}{\small $^\dagger$Fine structure of the line was unresolvable so only one gaussian was used.}
    \end{tabular*}
    \label{tab:perseus_lines}
\end{table*}

\begin{table*}
    \centering
    \caption{Fit results of emission line fluxes and line widths with a fixed \texttt{bapec} model described in \ref{subsec:Line Ratios}. Emission lines with all parameters tied to other lines are not included. Some lines were not detectable in C1 so they were either omitted or the doublet modeled as a single gaussian.}
    \renewcommand{\arraystretch}{1.25}
    \begin{tabular*}{\linewidth}{@{\extracolsep{\fill}} lcccc}
        \hline
        Line name & \multicolumn{2}{c}{C0} & \multicolumn{2}{c}{C1} \\
        \cline{2-5}
         & $\sigma$ (eV) & Flux ($\times10^{-5}$ cts/cm$^2$/s) & $\sigma$ (eV) & Flux ($\times10^{-5}$ cts/cm$^2$/s)\\
        \hline\hline
        Si {\scriptsize XIII} w & $0.20^{+1.01}_{-0.20}$ & $13.67^{+5.47}_{-4.87}$ & -- & -- \\
        
        Si {\scriptsize XIV} Ly$\alpha_1$ & $1.63^{+0.35}_{-0.35}$ & $38.49^{+3.57}_{-3.42}$ & $3.27^{+1.20}_{-1.06}$ & $84.50^{+22.84}_{-20.55}$ \\
        
        Si {\scriptsize XIV} Ly$\beta_1$ & $1.23^{+0.72}_{-0.95}$ & $5.93^{+1.14}_{-1.05}$ & $4.18^{+2.39}_{-1.41}$ & $20.27^{+8.97}_{-7.95}$ \\
        
        S {\scriptsize XV} w & $1.83^{+0.86}_{-0.88}$ & $5.30^{+1.37}_{-1.26}$ & $2.62^{+3.08}_{-2.62}$ & $12.78^{+8.68}_{-7.22}$ \\
        
        S {\scriptsize XVI} Ly$\alpha_1$ & $1.95^{+0..22}_{-0.23}$ & $21.41^{+1.07}_{-1.05}$ & $1.08^{+0.62}_{-1.08}$ & $43.84^{+5.85}_{-5.48}$ \\
        \hline
        S {\scriptsize XVI} Ly${\beta_1}^\ddagger$ & $1.91^{+0.45}_{-0.43}$ & $3.25^{+0.41}_{-0.39}$ & $1.15^{+2.67}_{-1.15}$ & $7.09^{+3.95}_{-3.03}$ \\
        
        S {\scriptsize XVI} Ly${\gamma_1}^\ddagger$ & $0.20^{+1.16}_{-0.20}$ & $0.71^{+0.23}_{-0.22}$ & $1.07^\dagger$ & $1.60^{+2.33}_{-1.60}$ \\
        
        Ar {\scriptsize XVII} w & $3.11^{+1.87}_{-1.30}$ & $3.67^{+0.98}_{-0.81}$ & $6.01^\dagger$ & $8.01^{+3.07}_{-4.58}$ \\
        
        Ar {\scriptsize XVIII} Ly$\alpha_1$ & $3.51^{+0.61}_{-0.59}$ & $5.00^{+0.46}_{-0.45}$ & $1.09^{+1.31}_{-1.06}$ & $12.55^{+2.50}_{-2.26}$ \\
        
        Ar {\scriptsize XVIII} Ly$\beta_1$ & $0.96^{+1.20}_{-0.96}$ & $0.50^{+0.18}_{-0.17}$ & $2.57^{+2.23}_{-1.40}$ & $1.63^{+1.28}_{-1.17}$ \\
        
        Ca {\scriptsize XIX} w & $3.24^{+0.49}_{-0.46}$ & $3.90^{+0.42}_{-0.40}$ & $6.48^{+2.34}_{-1.93}$ & $13.92^{+3.62}_{-3.36}$ \\
        
        Ca {\scriptsize XIX} He$\beta_1$ & $0.68^{+1.35}_{-0.68}$ & $0.44^{+0.19}_{-0.17}$ & -- & -- \\
        
        Ca {\scriptsize XX} Ly$\alpha_1$ & $3.67^{+0.79}_{-0.71}$ & $2.20^{+0.27}_{-0.26}$ & $3.98^{+2.22}_{-1.72}$ & $9.97^{+2.10}_{-1.89}$ \\
        \hline
        Fe {\scriptsize XXV} z & $3.71^{+0.18}_{-0.17}$ & $10.65^{+0.38}_{-0.37}$ & $3.66^{+0.37}_{-0.35}$ & $33.81^{+2.43}_{-2.32}$ \\
        
        Fe {\scriptsize XXV} w & $4.31^{+0.10}_{-0.09}$ & $28.19^{+0.55}_{-0.54}$ & $3.84^{+0.17}_{-0.17}$ & $88.88^{+3.42}_{-3.40}$ \\
        
        Fe {\scriptsize XXV} He$\beta_1$ & $4.93^{+0.44}_{-0.41}$ & $4.11^{+0.30}_{-0.29}$ & $3.86^{+0.76}_{-0.74}$ & $14.07^{+1.93}_{-1.79}$ \\
        
        Fe {\scriptsize XXV} He$\beta_2$ & $4.93^{+0.44}_{-0.41}$ & $0.72^{+0.21}_{-0.22}$ & $3.86^{+0.76}_{-0.74}$ & $2.47^{+1.27}_{-1.29}$ \\
        
        Fe {\scriptsize XXV} He$\gamma_1$ & $5.90^{+1.15}_{-0.99}$ & $1.47^{+0.21}_{-0.20}$ & $3.14^{+1.17}_{-0.91}$ & $4.75^{+1.15}_{-1.04}$ \\
        
        Fe {\scriptsize XXV} He$\delta_1$ & $4.40^{+1.32}_{-1.11}$ & $0.70^{+0.16}_{-0.15}$ & -- & -- \\
        
        Fe {\scriptsize XXV} He$\epsilon_1$ & $9.69^{+6.39}_{-3.93}$ & $0.37^{+0.21}_{-0.18}$ & $11.70^{+4.92}_{-3.67}$ & $3.02^{+1.41}_{-1.28}$ \\
        
        Fe {\scriptsize XXVI} Ly$\alpha_1$ & $4.16^{+0.32}_{-0.31}$ & $3.17^{+0.23}_{-0.22}$ & $2.47^{+0.36}_{-0.34}$ & $13.68^{+1.52}_{-1.45}$ \\
        
        Fe {\scriptsize XXVI} Ly$\alpha_2$ & $4.16^{+0.32}_{-0.31}$ & $1.73^{+0.18}_{-0.18}$ & $2.47^{+0.36}_{-0.34}$ & $8.68^{+1.28}_{-1.21}$ \\
        
        Fe {\scriptsize XXVI} Ly$\beta_1$ & $8.08^{+3.49}_{-3.38}$ & $0.50^{+0.15}_{-0.14}$ & $4.29^{+4.19}_{-4.29}$ & $2.16^{+0.91}_{-0.79}$ \\
        
        Ni {\scriptsize XXVII} w & $5.51^{+1.20}_{-0.97}$ & $1.29^{+0.21}_{-0.19}$ & $7.03^{+12.21}_{-2.83}$ & $2.78^{+2.04}_{-1.18}$ \\
        \hline
        \multicolumn{5}{l}{\small $^\dagger$Value is an upper bound.}\\
        \multicolumn{5}{l}{\small $^\ddagger$Doublet replaced by single gaussian in C1 due to insufficient signal.}
    \end{tabular*}
    \label{tab:perseus_lines_results}
\end{table*}

\texttt{btapec} models were attempted to separately solve for continuum and emission line temperatures following the results of Figure \ref{fig:line_ratios}.   However, as discussed in Section \ref{Discussion:lineratios}, the model failed to distinguish between the emission line and continuum temperatures in either pointing.

\subsection{Multiple Temperature Models}\label{subsec:multi-T}

To examine the temperature structure in greater detail the footprint was divided into 5 sub-FOV regions shown in Figure \ref{fig:pointings}. These regions were chosen to optimize count statistics and spatial resolution in the C0 \& C1 pointings. Simultaneous fits of all regions were executed using spatial-spectral mixing to account for photon contamination over the 1.8--10 keV band. The spectra were binned to a minimum of one count per bin. One (\texttt{bvvapec}), two- (\texttt{bvvapec+bvvapec}), and temperature distribution (\texttt{bvvgadem}) models were used.  The temperature distribution model assumed a gaussian distribution of temperatures.

The Fe {\scriptsize XXV} He$\alpha$ w line was removed in regions 1 and 2 and replaced with a redshifted gaussian to account for resonant scattering. A single gaussian is included in each region with the rest frame energy ($E_0=6700.4$ eV) fixed.  When a one temperature model per region is adopted, the redshift of the line is tied to the model. Otherwise, the redshift is free to allow for velocity variance associated with multiple temperature phases. A power law component is included in regions 1--3 to account for the AGN. The power law flux over 2--10 keV was fixed in accordance with best-fit values found in the \citetalias{XRISM_Perseus} as $\mathrm{Flux}_{2-10\,\mathrm{keV}}=31\times10^{-12}$ erg cm$^{-2}$. The photon index, $\Gamma$, was free to vary. The abundances of the two-temperature model are tied between the two components.

The best temperature model was identified using the Bayesian Information Criterion (BIC) defined as,
\begin{equation}
    \mathrm{BIC}=k\ln{n}-2\ln{\hat{L}}.
\end{equation}
Here $k$ is the number of model parameters, $n$ is the number of data points, and $\hat{L}$ is the likelihood. The C-statistic is equivalent to the second term \citep{Cash}. Therefore,
\begin{equation}
    \mathrm{BIC}=k\ln{n} + \text{C-stat}.
\end{equation}
The BIC penalizes overfitting by unnecessary parameters.  In general a $\Delta\mathrm{BIC}\geq10$ indicates the model with the lower BIC is preferred. The temperature distribution and two-temperature models have one and five more free parameters than one-temperature, respectively.

One temperature, two temperatures, and multiple temperature models were tested for the outer regions 3--5. Regions 3--5 could not constrain a second temperature component, nor did a temperature distribution model improve the quality of the fit in regions 3--5. Therefore, the outer regions are best described by a single temperature which is consistent with observations \citep[e.g.][]{FabianPerseus2006}. The inner two regions are more complex as described in sub-sections \ref{subsec:combined_central_region} \& \ref{subsec:separate_central_region} below.

\subsubsection{Combined central region}\label{subsec:combined_central_region}

The emission line ratios analysis indicated that multiple temperatures are difficult to discern, even in the central pointings. Therefore, regions 1 and 2 were first combined (henceforth region 1+2) to maximize signal in the central region at the expense of lower spatial resolution. One-, two-, and broad temperature distribution models were tested. The two-temperature model significantly improved the statistics over one-temperature while the temperature distribution model did not. We find $\Delta\mathrm{BIC}=27$ statistical significance for two temperatures over one temperature in the combined region.

\begin{figure}[ht]
    \centering
    \includegraphics[width=\columnwidth]{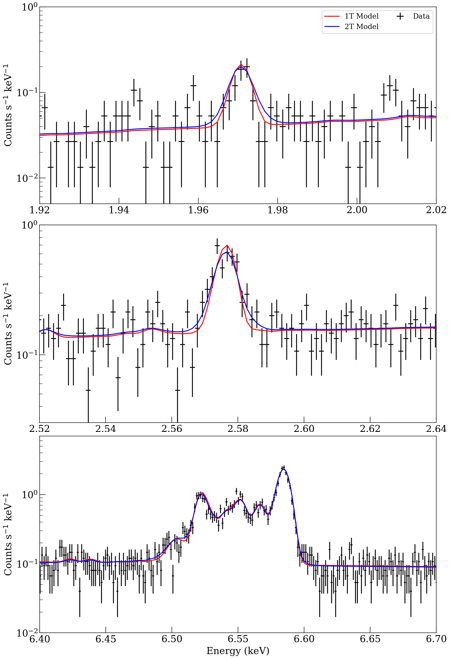}
    \caption{Comparison of the region 2 spectrum between the model fit with two temperatures in region 2 (blue) and a one-temperature model (red) for Si {\scriptsize XIV} Ly$\alpha$ (top), S {\scriptsize XVI} Ly$\alpha$ (middle), and Fe {\scriptsize XXV} He$\alpha$ (bottom). The spectrum is re-binned for plotting purposes only. Contributions from the power law, NXB, Fe {\scriptsize XXV} He$\alpha$ w line gaussian, and scattered photons are not shown.}\label{fig:multiT_lines_comparison}
\end{figure}

The best-fit results are provided in Table \ref{tab:SSM_combined_fit}. With the best-fit model, a second component which features a large ($316^{+45}_{-37}$ km\,s$^{-1}$) velocity dispersion and cooler temperature ($2.43^{+0.14}_{-0.15}$ keV) is detected. The hotter component tends to fit the mean of the temperature, redshift, and velocity dispersion profiles across regions 1 and 2. Close inspection of the emission lines verified that the two-temperature model, in addition to small adjustments for temperature and redshift variation between regions 1 and 2, required a cooler and broader component to properly fit broad features of prominent emission lines (e.g. Figure \ref{fig:multiT_lines_comparison}).

\begin{table*}[ht]
    \centering
    \caption{Best fit parameters of all sub-FOV regions, with regions 1 and 2 combined into a single region and modeled with two temperatures, with SSM from 1.8--10 keV. Details of each region model are provided in sections \ref{subsec:multi-T} \& \ref{subsec:combined_central_region}. Some abundances were not constrainable in the outer regions, so those parameters were frozen at solar values. The power law photon index best-fit value is $\Gamma=1.83^{+0.07}_{-0.06}$. Fit C-stat/dof = 121380/131139.}
    \renewcommand{\arraystretch}{1.25}
    \begin{tabular*}{\linewidth}{@{\extracolsep{\fill}} lcccc}
    \hline
    Parameter & Combined Region 1 + 2 & Region 3 & Region 4 & Region 5 \\
    \hline
    Radius (arcmin) & 0 -- 3 & 3 -- 5 & 5 -- 7 & 7 -- 11.85 \\
    
    $k_BT$ (keV) & $4.43^{+0.10}_{-0.09}$ & $5.31^{+0.12}_{-0.12}$ & $5.88^{+0.18}_{-0.18}$ & $7.42^{+0.26}_{-0.26}$ \\

    Si abund. & $0.74^{+0.07}_{-0.06}$ & $0.56^{+0.31}_{-0.27}$ & $0.57^{+0.52}_{-0.43}$ & -- \\
    
    S abund. & $0.70^{+0.04}_{-0.04}$ & $0.51^{+0.16}_{-0.15}$ & $0.88^{+0.29}_{-0.26}$ & $0.45^{+0.27}_{-0.24}$ \\
    
    Ar abund. & $0.62^{+0.05}_{-0.05}$ & $0.74^{+0.24}_{-0.22}$ & $0.93^{+0.40}_{-0.37}$ & $0.37^{+0.43}_{-0.37}$ \\
    
    Ca abund. & $0.65^{+0.04}_{-0.04}$ & $0.66^{+0.19}_{-0.18}$ & -- & -- \\
    
    Fe abund. & $0.71^{+0.01}_{-0.01}$ & $0.50^{+0.02}_{-0.02}$ & $0.53^{+0.03}_{-0.03}$ & $0.36^{+0.03}_{-0.03}$ \\

    Ni abund. & $0.71^{+0.06}_{-0.06}$ & $0.45^{+0.17}_{-0.16}$ & $0.98^{+0.30}_{-0.28}$ & $0.29^{+0.24}_{-0.22}$ \\
    
    z ($\times10^{-2}$) & $1.761^{+0.002}_{-0.002}$ & $1.700^{+0.003}_{-0.003}$ & $1.700^{+0.006}_{-0.006}$ & $1.776^{+0.008}_{-0.008}$ \\
    
    $\sigma$ (km\,s$^{-1}$) & $138^{+8}_{-8}$ & $84^{+13}_{-14}$ & $183^{+18}_{-17}$ & $202^{+25}_{-23}$ \\
    
    Norm ($\times10^{-2}$) & $22.11^{+0.16}_{-0.18}$ & $18.17^{+0.35}_{-0.35}$ & $3.86^{+0.09}_{-0.09}$ & $3.84^{+0.07}_{-0.07}$ \\
    
    ${k_BT}_2$ (keV) & $2.43^{+0.15}_{-0.14}$ & -- & -- & -- \\
    
    z$_2$ ($\times10^{-2}$) & $1.787^{+0.011}_{-0.010}$ & -- & -- & -- \\
    
    $\sigma_2$ (km\,s$^{-1}$) & $316^{+45}_{-36}$ & -- & -- & -- \\
    
    Norm$_2$ ($\times10^{-2}$) & $11.68^{+1.89}_{-1.73}$ & -- & -- & -- \\
    \hline
    \end{tabular*}
    \label{tab:SSM_combined_fit}
\end{table*}

\subsubsection{Finer spatial binning}\label{subsec:separate_central_region}

The analysis of the combined region indicates a second component in the inner $\sim60$ kpc. The multiple temperature analysis with the central 2 regions separated (regions 1 and 2) was performed to examine finer spatial detail.

Combinations of one-, two-, and temperature distribution models were tested for regions 1 and 2. When using a one-temperature for one region and two-temperature for the other in either combination provided significant statistical improvement to the fit over a one-temperature model with $\Delta\mathrm{BIC}=12$ for two temperatures in region 1 and $\Delta\mathrm{BIC}=78$ for two temperatures in region 2. The statistics strongly preferred two temperatures in region 2 over two temperatures in region 1 or the combined region. In both cases a cooler component with high ($\geq290$ km\,s$^{-1}$) velocity dispersion was indicated. The use of temperature distribution in either or both regions provided no significant statistical or qualitative improvement to the fit. The best fit for the model with two temperatures in region 2 is provided in Table \ref{tab:SSM_full_fit}, and the best fit for the model with two temperatures in region 1 is provided in Table \ref{tab:SSM_full_fit_2Treg1}.

Regions 1 and 2 were fitted with two temperatures simultaneously. However, both second temperature components were unconstrained in region 1 due to noise in the soft band.  This component already had preference for very low ($k_BT<2$ keV) temperature in the 2T + 1T case (Table \ref{tab:SSM_full_fit_2Treg1}), which propagated issues to the model in region 2. Stricter limits were imposed on the parameters of the models in regions 1 and 2 (e.g. $k_BT\leq6$ keV, $z\sim0.016-0.019$, $\sigma\leq500$ km s$^{-1}$). However, this caused most or all parameters of the second component of both regions to pin to the imposed limits. Tying the redshifts, velocity dispersions, or both between the two components in region 1 in addition to stricter parameter limits were unsuccessful.  Therefore, we were unable to successfully fit a model with two temperatures in both regions 1 and 2 simultaneously.

\begin{table*}
    \centering
    \caption{Best fit parameters of all sub-FOV regions with SSM from 1.8--10 keV, with two temperatures in region 2. Details of each region model are provided in sections \ref{subsec:multi-T} \& \ref{subsec:separate_central_region}. Some abundances were not constrainable in the outer regions, so those parameters were frozen at solar values. The power law photon index best-fit value is $\Gamma=1.68^{+0.05}_{-0.09}$. Fit C-stat/dof = 156587/163923.}
    \renewcommand{\arraystretch}{1.25}
    \begin{tabular*}{\linewidth}{@{\extracolsep{\fill}} lccccc}
    \hline
    Parameter & Region 1 & Region 2 & Region 3 & Region 4 & Region 5 \\
    \hline
    Radius (arcmin) & 0 -- 0.9 & 0.9 -- 3 & 3 -- 5 & 5 -- 7 & 7 -- 11.85 \\
    
    $k_BT$ (keV) & $3.49^{+0.11}_{-0.19}$ & $4.85^{+0.10}_{-0.09}$ & $5.29^{+0.12}_{-0.12}$ & $5.88^{+0.18}_{-0.17}$ & $7.42^{+0.25}_{-0.26}$\\
    
    Si abund. & $1.02^{+0.18}_{-0.17}$ & $0.55^{+0.10}_{-0.09}$ & $0.61^{+0.32}_{-0.28}$ & $0.56^{+0.53}_{-0.44}$ & --\\
    
    S abund. & $0.86^{+0.10}_{-0.14}$ & $0.55^{+0.06}_{-0.05}$ & $0.55^{+0.17}_{-0.15}$ & $0.86^{+0.29}_{-0.26}$ & $0.45^{+0.27}_{-0.24}$\\
    
    Ar abund. & $0.60^{+0.12}_{-0.11}$ & $0.60^{+0.08}_{-0.08}$ & $0.75^{+0.24}_{-0.23}$ & $0.92^{+0.40}_{-0.36}$ & $0.37^{+0.43}_{-0.37}$\\
    
    Ca abund. & $0.57^{+0.10}_{-0.10}$ & $0.66^{+0.08}_{-0.07}$ & $0.66^{+0.19}_{-0.18}$ & -- & --\\
    
    Fe abund. & $0.68^{+0.03}_{-0.03}$ & $0.74^{+0.02}_{-0.02}$ & $0.48^{+0.02}_{-0.02}$ & $0.54^{+0.03}_{-0.03}$ & $0.36^{+0.03}_{-0.03}$\\
    
    Ni abund. & $0.51^{+0.18}_{-0.18}$ & $0.85^{+0.12}_{-0.11}$ & $0.44^{+0.17}_{-0.16}$ & $0.97^{+0.29}_{-0.27}$ & $0.29^{+0.24}_{-0.22}$\\
    
    z ($\times10^{-2}$) & $1.792^{+0.003}_{-0.003}$ & $1.749^{+0.002}_{-0.003}$ & $1.700^{+0.003}_{-0.003}$ & $1.700^{+0.005}_{-0.005}$ & $1.776^{+0.008}_{-0.008}$\\
    
    $\sigma$ (km\,s$^{-1}$) & $141^{+13}_{-13}$ & $111^{+9}_{-9}$ & $91^{+14}_{-14}$ & $179^{+18}_{-17}$ & $203^{+25}_{-23}$\\
    
    Norm ($\times10^{-2}$) & $7.41^{+0.47}_{-0.24}$ & $15.39^{+.70}_{-0.91}$ & $17.97^{+0.35}_{-0.35}$ & $3.87^{+0.09}_{-0.09}$ & $3.84^{+0.07}_{-0.07}$\\
    
    ${k_BT}_2$ (keV) & -- & $2.30^{+0.11}_{-0.08}$ & -- & -- & --\\
    
    z$_2$ ($\times10^{-2}$) & -- & $1.744^{+0.016}_{-0.016}$ & -- & -- & --\\
    
    $\sigma_2$ (km\,s$^{-1}$) & -- & $428^{+64}_{-54}$ & -- & -- & --\\
    
    Norm$_2$ ($\times10^{-2}$) & -- & $12.78^{+1.03}_{-0.90}$ & -- & -- & --\\
    
    \hline
    \end{tabular*}
    \label{tab:SSM_full_fit}
\end{table*}

\begin{table*}
    \centering
    \caption{Best fit parameters of all sub-FOV regions with SSM from 1.8--10 keV, with two temperatures in region 1. Details of each region model are provided in sections \ref{subsec:multi-T} \& \ref{subsec:separate_central_region}. Some abundances were not constrainable in the outer regions, so those parameters were frozen at solar values. The power law photon index best-fit value is $\Gamma=1.60^{+0.05}_{-0.05}$. Fit C-stat/dof = 156654/163923.}
    \renewcommand{\arraystretch}{1.25}
    \begin{tabular*}{\linewidth}{@{\extracolsep{\fill}} lccccc}
    \hline
    Parameter & Region 1 & Region 2 & Region 3 & Region 4 & Region 5 \\
    \hline
    Radius (arcmin) & 0 -- 0.9 & 0.9 -- 3 & 3 -- 5 & 5 -- 7 & 7 -- 11.85 \\
    
    $k_BT$ (keV) & $4.27^{+0.31}_{-0.20}$ & $3.96^{+0.04}_{-0.04}$ & $5.28^{+0.12}_{-0.12}$ & $5.89^{+0.18}_{-0.18}$ & $7.42^{+0.26}_{-0.26}$\\
    
    Si abund. & $0.59^{+0.12}_{-0.10}$ & $0.72^{+0.13}_{-0.12}$ & $0.61^{+0.32}_{-0.28}$ & $0.56^{+0.53}_{-0.44}$ & --\\
    
    S abund. & $0.46^{+0.08}_{-0.07}$ & $0.79^{+0.07}_{-0.07}$ & $0.48^{+0.16}_{-0.15}$ & $0.89^{+0.29}_{-0.26}$ & $0.45^{+0.27}_{-0.24}$\\
    
    Ar abund. & $0.48^{+0.10}_{-0.09}$ & $0.67^{+0.08}_{-0.08}$ & $0.70^{+0.24}_{-0.22}$ & $0.94^{+0.40}_{-0.37}$ & $0.37^{+0.43}_{-0.37}$\\
    
    Ca abund. & $0.65^{+0.10}_{-0.10}$ & $0.58^{+0.07}_{-0.07}$ & $0.65^{+0.19}_{-0.18}$ & -- & --\\
    
    Fe abund. & $0.74^{+0.03}_{-0.03}$ & $0.69^{+0.02}_{-0.02}$ & $0.48^{+0.02}_{-0.02}$ & $0.54^{+0.03}_{-0.03}$ & $0.36^{+0.03}_{-0.03}$\\
    
    Ni abund. & $0.59^{+0.21}_{-0.20}$ & $0.76^{+0.11}_{-0.10}$ & $0.44^{+0.17}_{-0.16}$ & $0.97^{+0.29}_{-0.27}$ & $0.29^{+0.24}_{-0.22}$\\
    
    z ($\times10^{-2}$) & $1.787^{+0.005}_{-0.005}$ & $1.749^{+0.002}_{-0.002}$ & $1.700^{+0.003}_{-0.003}$ & $1.700^{+0.005}_{-0.005}$ & $1.776^{+0.008}_{-0.008}$\\
    
    $\sigma$ (km\,s$^{-1}$) & $127^{+17}_{-19}$ & $143^{+9}_{-8}$ & $92^{+14}_{-14}$ & $179^{+18}_{-17}$ & $203^{+25}_{-23}$\\
    
    Norm ($\times10^{-2}$) & $4.39^{+0.49}_{-0.76}$ & $25.02^{+0.30}_{-0.30}$ & $18.02^{+0.35}_{-0.36}$ & $3.87^{+0.09}_{-0.09}$ & $3.84^{+0.07}_{-0.07}$\\
    
    ${k_BT}_2$ (keV) & $1.87^{+0.20}_{-0.13}$ & -- & -- & -- & --\\
    
    z$_2$ ($\times10^{-2}$) & $1.809^{+0.014}_{-0.014}$ & -- & -- & -- & --\\
    
    $\sigma_2$ (km\,s$^{-1}$) & $295^{+62}_{-51}$ & -- & -- & -- & --\\
    
    Norm$_2$ ($\times10^{-2}$) & $4.97^{+0.78}_{-0.68}$ & -- & -- & -- & --\\
    
    \hline
    \end{tabular*}
    \label{tab:SSM_full_fit_2Treg1}
\end{table*}

\subsubsection{Abundances}\label{subsec:abundances}
The two-temperature model also reports abundances for Si, S, Ar, Ca, Fe, and Ni (rows 3--8 of Tables \ref{tab:SSM_combined_fit}--\ref{tab:SSM_full_fit_2Treg1}). As the model with two temperatures in region 2 reports the highest statistical significance, we calculate the abundance ratios of this model (Table \ref{tab:SSM_full_fit}). Assuming that the abundances of each element in each region are mostly independent from each other, and that the errors are gaussian, we calculate the weighted mean and uncertainty of the ratio of each element with respect to Fe.
For the regions where Si and Ca were not constrained their ratios are not reported. The mean abundance ratios with respect to Fe are provided in Table \ref{tab:abundance_ratios}. These ratios are consistent with the Solar System.  Additionally, the abundance ratios are consistent with the more thorough abundance analysis in an upcoming paper. Although tying the abundances of the cooler and hotter components together in the two temperature models was necessary for successful spectral fitting, it is unknown whether the two gas components should have the same chemical composition.

\begin{table}[h]
    \centering
    \caption{Mean abundance ratios with respect to Fe.}
    \renewcommand{\arraystretch}{1.25}
    \begin{tabular}{ccccc}
    \hline
    Si/Fe & S/Fe & Ar/Fe & Ca/Fe & Ni/Fe \\
    $1.06^{+0.10}_{-0.10}$ & $0.89^{+0.07}_{-0.07}$ & $0.87^{+0.09}_{-0.09}$ & $0.90^{+0.09}_{-0.09}$ & $1.06^{+0.12}_{-0.12}$ \\
    \hline
    \end{tabular}
    \label{tab:abundance_ratios}
\end{table}

\subsection{Multiple Temperature Systematics}\label{subsec:systematics}

\subsubsection{The AGN component}\label{subsec:systematics:agn}
When both the power law normalization and photon index are allowed to vary, Xspec is unable to find parameter values consistent with observations unless strict constraints are imposed. Therefore the power law flux over 2--10 keV was held constant $\mathrm{Flux}_{2-10\,\mathrm{keV}}=31\times10^{-12}$ erg cm$^{-2}$ s at the best-fit value found in the \citetalias{XRISM_Perseus}. Depending on the model, this results in a best-fit value for the photon index of $\Gamma=1.60-1.83$, which is consistent with the \citetalias{XRISM_Perseus}. Fixing the power law photon index instead of the flux does not result in flux consistent with observations, as seen in the \citetalias{XRISM_Perseus}. At the very least, the second component temperature and velocity dispersion should be unaffected by differences in power law modeling.

The best-fit models were tested with two temperatures in regions 1 or 2 with the photon index fixed to $\Gamma=1.7$ and allowing the flux to vary. An example of this fit with two temperatures in region 2 is reported in Table \ref{tab:SSM_full_fit_appendix} in appendix \ref{appendix:AGN}. This fit results in a larger power law flux of $\mathrm{Flux}_{2-10\mathrm{\,keV}}=42^{+1}_{-1}\times10^{-12}$ erg cm$^{-2}$ s$^{-1}$, causing a temperature drop in region 1 along with an increase in Fe abundance. The cooler component temperature, abundances, and velocity dispersion are unaffected.  Any significant changes in parameter values in region 2 are consistent within $1\sigma$ with Table \ref{tab:SSM_full_fit}. The same exercise with the $\Delta\mathrm{BIC}$ was performed for variations of one- and two-temperature modeling for regions 1 and 2 with the power law photon index fixed instead of the flux. Similarly, the statistics preferred two temperatures in both regions.

\subsubsection{Atomic databases}
Models using multiple temperature components will be sensitive to finer details of the spectrum, especially emission lines. Differences in atomic databases can have a significant impact on multiple temperature models, as seen in \cite{Hitomi_Perseus_lineratios}. We tested the fits using the \texttt{SPEXACT} v3.07 and \texttt{AtomDB} v3.1.3 atomic databases instead of \texttt{AtomDB} v3.0.9. Fits using these atomic databases are consistent with the best-fit models reported in Tables \ref{tab:SSM_combined_fit}--\ref{tab:SSM_full_fit_2Treg1}.

\subsubsection{Resonant scattering}\label{subsec:resonant_scattering}
The Fe {\scriptsize XXV} He$\alpha$ w line is suppressed due to resonant scattering in the cluster core \citep{Perseus_Hitomi_ResonantScattering}, which impacts spectral fitting using standard CIE models. This issue was addressed by removing the line from the plasma model in regions 1 and 2 which are affected most by resonant scattering and modeling the line with a gaussian. A gaussian may not necessarily be the best option to fit this line, which could influence the multiple temperature result since it is the brightest line in the spectrum. We tested the best-fit model using a Voigt profile, as well as ignoring a narrow energy band around the line from the spectrum. In both cases the fits are consistent with Tables \ref{tab:SSM_combined_fit}--\ref{tab:SSM_full_fit_2Treg1}.

Only a single gaussian is used in regions 1 and 2 in order to account for resonant scattering. For the two-temperature models the redshift of the gaussian is allowed to vary for deviations due to multiple temperatures, however it is ideal to use two gaussians to properly account for the scattered line from each temperature component. The two-temperature models were fit to the spectra with two gaussians to account for resonant scattering. Tying the relative fluxes of the gaussians to the relative fluxes of the two temperature components was a necessary constraint for successful model fitting. These fits are consistent with Tables \ref{tab:SSM_combined_fit} and \ref{tab:SSM_full_fit}. However, we could not successfully model two temperatures in region 1 (Table \ref{tab:SSM_full_fit_2Treg1}), with the additional gaussian component. This is due to insufficient count statistics to constrain the complex model in this region, as seen when trying to fit two temperatures in both regions 1 \& 2 simultaneously (sub-section \ref{subsec:separate_central_region}). Therefore, the reported fits are those with only a single gaussian for consistency across all models.

\subsubsection{Spatial-spectral mixing and radial binning}
The use of spatial-spectral mixing model compensates for contributions to the spectral model in a specific region from other regions in the sky scattered in by the broad point spread function.
However, the spectral fitting may not model the relative fluxes of multiple regions or components consistent with observations in favor of a statistical local minimum, as was seen with the AGN component in the \citetalias{XRISM_Perseus}. Therefore, we must verify that the relative fluxes of each region, especially those with contributions from multiple components (e.g. multiple temperatures, the AGN component), from the fits are consistent with observations.

The approach applied here to constrain the relative fluxes of regions 1 and 2 for the multiple temperature fits is the same as \citetalias{XRISM_Perseus}.  The Chandra observations of Perseus (ObsIDs: 3209, 4952, 11714) were reprocessed with CIAO 4.17 as in \citet{CIAO}. Spectra from regions 1 and 2 were extracted from each Chandra observation with \texttt{specextract}, removing a $\sim9$ arcsec diameter circle from the center of region 1 to remove contributions from the AGN point source. Each region was fit over 4--7.3 keV with a \texttt{TBabs*vapec} model with $\text{N}_{\text{H}} = 1.38\times10^{21}$ cm$^{-2}$ fixed and all abundances tied to Fe. We then measured the flux over 4--6 keV, which is dominated by continuum emission, for each region. We then constrained the multiple temperature XRISM fit by fixing the relative fluxes between regions 1 and 2. The fits using this constraint is consistent with that in Tables \ref{tab:SSM_full_fit}--\ref{tab:SSM_full_fit_2Treg1}.

The sub-FOV regions are binned radially outward from the center of the cluster. These regions are different than the main profile reported in the \citetalias{XRISM_Perseus}. The immediate effect of this is seen in the velocity dispersion profile in Fig. \ref{fig:multiT_profiles}, where the velocity jump at $\sim7$ arcmin is not detected. Despite this difference, the trends of the temperature and redshift are unaffected and the differences in velocity dispersion between the data points of the relative outer regions are minimal as seen in the \citetalias{XRISM_Perseus}. Additionally, the cooler component is detected only in the inner regions. 
The outer regions are included in SSM to more accurately fit region 3, which is less affected by photons from regions 4 and 5 than regions 1 and 2. While this approach introduces additional, albeit minor, uncertainty the focus of this work is on the multiple temperature structure of the inner regions which is largely unaffected by this difference.

\subsubsection{Energy scale uncertainty below 5.4 keV}
Due to XRISM/Resolve’s gate valve closure, the energy scale below 5.4 keV, the lowest energy accessible on-board calibration source, has higher measurement uncertainty. Guidance from the Instrument Team cautions that an energy scale uncertainty of $\pm1$ eV below 5.4 keV may apply to some observations. This guidance is based on measurements of the instrumental lines of Si K and Al K. An energy scale offset may result in a false redshift or an increased velocity dispersion which can have nonlinear effects on spectral modeling.

The multiple temperature model is sensitive to the softer energies since the cooler component is sensitive to the Si and S emission lines below $\sim3$ keV. If the model fit of a region that is confidently fit with a single temperature finds varying redshift or velocity dispersion over different narrow bands, it may indicate an energy scale offset. We performed single temperature narrow band fits of 2--4, 4--6, and 6--7 keV for regions 3, 4, and 5 without SSM. Unfortunately, low photon counts in these regions provide poor constraints on redshift and poor or no constraints on velocity dispersion over narrow bands that do not include the Fe {\scriptsize XXV} emission lines. The lack of precise constraints prevents us from measuring or ruling out an energy scale offset.

If we assume the existence of an energy scale offset below 5.4 keV, the $\pm1$ eV uncertainty translates to an uncertainty of $\sim 150$ km s$^{-1}$ at 2 keV. Assuming this was measured as purely additional velocity dispersion, then the uncertainty affects the measured velocity dispersions of the cooler component in quadrature. Therefore in the worst case the cooler component velocity dispersions are inflated by up to $\sim 30-40$ km s$^{-1}$. Although this introduces additional uncertainty, this is only at most a $\sim10$\% effect, which does not rule out the high velocity dispersion of the cooler component. However, this uncertainty can affect both redshift and velocity dispersion, so the true effects on the multiple temperature modeling may be more complex.

\section{Discussion}\label{Discussion}

\subsection{Emission Line Ratios}\label{Discussion:lineratios}
\begin{figure*}
    \centering
    \includegraphics[width=\linewidth]{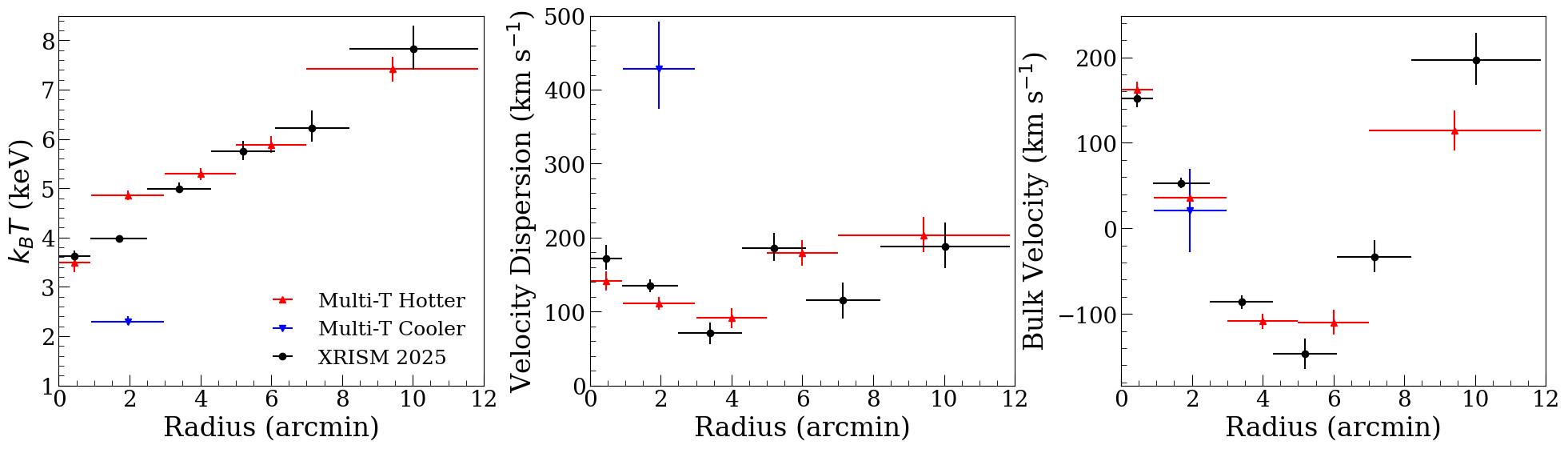}
    \caption{Best-fit model, with two temperatures in region 2, measured radial profiles of temperature (left), velocity dispersion (middle), and bulk velocity (right) (Table \ref{tab:SSM_full_fit}). The black data points represent the values from the \citetalias{XRISM_Perseus}. The red data points represent the values from the single-temperature or hotter component of the two-temperature model. The blue data points represent the values from the cooler component of the two-temperature model. The bulk velocities are calculated as $v_\mathrm{bulk}=c(z-z_\mathrm{BCG})/(1+z_\mathrm{BCG})$, relative to the BCG redshift $z_\mathrm{BCG}=0.017284$, with heliocentric correction. The redshift values, $z$, are from the Xspec best-fit model.}\label{fig:multiT_profiles}
\end{figure*}

In the central pointing, the scatter in the excitation temperature measured from the emission lines of Si, S Ar, and Ca (Fig. \ref{fig:line_ratios}) across 0--2 keV may reflect multiple temperature gas, Fe measurements provide constraints on the hottest gas. The Fe measurements for the C1 pointing are consistent with both the Hitomi results and the C0 pointing, albeit with larger uncertainties. Measurements of Si, S, and Ar are too noisy to provide constraints. For the ionization temperatures, only Si indicates a potential cooler component. S, Ar, and Ca are consistent with the Hitomi results, as are the C0 flux ratios.  However, there is some preference for hotter gas in C1. The temperature measurements from Fe for both C0 and C1 are consistent with the temperatures measured in the \citetalias{XRISM_Perseus} (Fig. \ref{fig:multiT_profiles}), with the exception of He$\epsilon$ for C1 which is a largely uncertain measurement.

The higher temperature gas, measured via multiple temperature modeling, is consistent with the emission line ratio measurements of Fe (Fig. \ref{fig:line_ratios}), while the cooler gas temperatures are roughly consistent with excitation temperature measurements of Si--Ca and ionization temperature measurements of Si.

Both pointings have shorter integration times than Hitomi which explains the larger error bars on all flux ratio measurements.  It also explains why we were unable to obtain separate constraints on the continuum and emission temperatures with the \texttt{btapec} model. Due to the large uncertainties, the flux ratios are consistent with a single temperature CIE gas.

\subsection{Multiple Temperature Modeling}
\subsubsection{Analysis summary}
The multiple temperature analysis divides the C0 and C1 pointings into three sub-FOV regions, two of which show a preference for multiple temperatures. The results of the multiple temperature modeling (section \ref{subsec:multi-T}) are:
\begin{itemize}
    \setlength\itemsep{0em}
    \item the outer three regions are best described with a single temperature model, which is consistent with previous observations.
    \item the emission line ratio analysis indicates that deviations due to CIE would be challenging to model, so a larger central region (region 1+2) was modeled combining the full C0 pointing and part of the C1 pointing. A second component was indicated with a cooler temperature of $2.43^{+0.14}_{-0.15}$ keV and higher velocity dispersion of $316^{+45}_{-37}$ km s$^{-1}$, with a statistical significance of $\Delta\mathrm{BIC}=27$ over one temperature.
    \item the central region was divided into two smaller regions to examine the spectra with finer spatial detail. Significant statistical improvement to the fit over one temperature was found when modeling one region with two temperatures and the other with one temperature.  Both indicate a second cooler component with higher velocity dispersion. The statistics strongly preferred two temperatures in region 2. Modeling both regions with two temperatures simultaneously was unsuccessful.
    \item when modeling two temperatures in region 1, the second component has a temperature of $1.87^{+0.20}_{-0.13}$ keV and velocity dispersion of $292^{+62}_{-51}$ km s$^{-1}$.  The statistical significance is found to be $\Delta\mathrm{BIC}=12$ over one temperature. Modeling two temperatures in region 2 yields a second component with a temperature of $2.30^{+0.10}_{-0.08}$ keV and velocity dispersion of $429^{+64}_{-54}$ km s$^{-1}$. Its statistical significance is $\Delta\mathrm{BIC}=78$ over one temperature.
\end{itemize}

Regardless of model choice, a second component with cooler temperature and higher velocity dispersion is indicated in the central $\sim60$ kpc with high statistical significance. This component, especially the high velocity dispersion, is robust against systematic effects from modeling and instrument uncertainty (see section \ref{subsec:systematics}). Additionally, the second component has a qualitative impact on the spectral fitting, seen in Figure \ref{fig:multiT_lines_comparison} and described in detail in the next section.

\begin{table}[h]
    \centering
    \caption{Model BIC significance summary. For each model, the region where two temperatures are used instead of one temperature is highlighted in bold, with reference to the table in which the fit is reported. The $\Delta\mathrm{BIC}$ represents the statistical significance of the model compared to the model fit with all regions as one temperature.}
    \renewcommand{\arraystretch}{1.25}
    \begin{tabular}{lcc}
    \hline
    \multicolumn{1}{c}{Model} & Table & $\Delta\mathrm{BIC}$ \\
    {\bf 2T} -- 1T -- 1T -- 1T & \ref{tab:SSM_combined_fit} & 27 \\
    1T -- {\bf 2T} -- 1T -- 1T -- 1T & \ref{tab:SSM_full_fit} & 78 \\
    {\bf 2T} -- 1T -- 1T -- 1T -- 1T & \ref{tab:SSM_full_fit_2Treg1} & 12 \\
    \hline
    \end{tabular}
    \label{tab:BIC_summary}
\end{table}

\subsubsection{Emission line structure}

Looking more closely at Si {\scriptsize XIV} \& S {\scriptsize XVI} Ly$\alpha$ and Fe {\scriptsize XXV} He$\alpha$ (Fig. \ref{fig:multiT_lines_comparison}), we see that the cooler component is dominant for the Si and S emission lines, and tends to be responsible for the wider broadening of the wings of most emission lines in the spectrum. The hot component is dominant for the other emission lines and tends to be responsible for the peaks of the lines. The impact of the cooler component is most clearly seen in S {\scriptsize XVI} Ly$\alpha$, capturing the broader wings of the line while not overshooting the peak. The two-temperature model prefers a broader width for Si {\scriptsize XIV} Ly$\alpha$, although it is unclear whether this is a better fit. The Fe {\scriptsize XXV} He$\alpha$ z line and wide wing at $\sim6.60$ keV are better fit by the two-temperature model, however it is unclear if the broader wing at $\sim6.50$ keV is a better fit. The Fe {\scriptsize XXV} He$\alpha$ y line is not well fit by either model, which is seen in other objects \citep[e.g.][]{XRISM_2025_A2029}. Although the wide wing of the Fe {\scriptsize XXV} He$\alpha$ w line appears to be better fit by the 2-T model, the gaussian component included to account for resonant scattering only accounts for the distortion of the line flux, and does not account for distortion in the line shape. Taking into account the systematics and modeling complexities discussed in section \ref{subsec:resonant_scattering}, it is unclear whether or not the two-temperature model fits the wing at $\sim6.60$ keV better.

\begin{figure*}
    \centering
    \includegraphics[width=\linewidth]{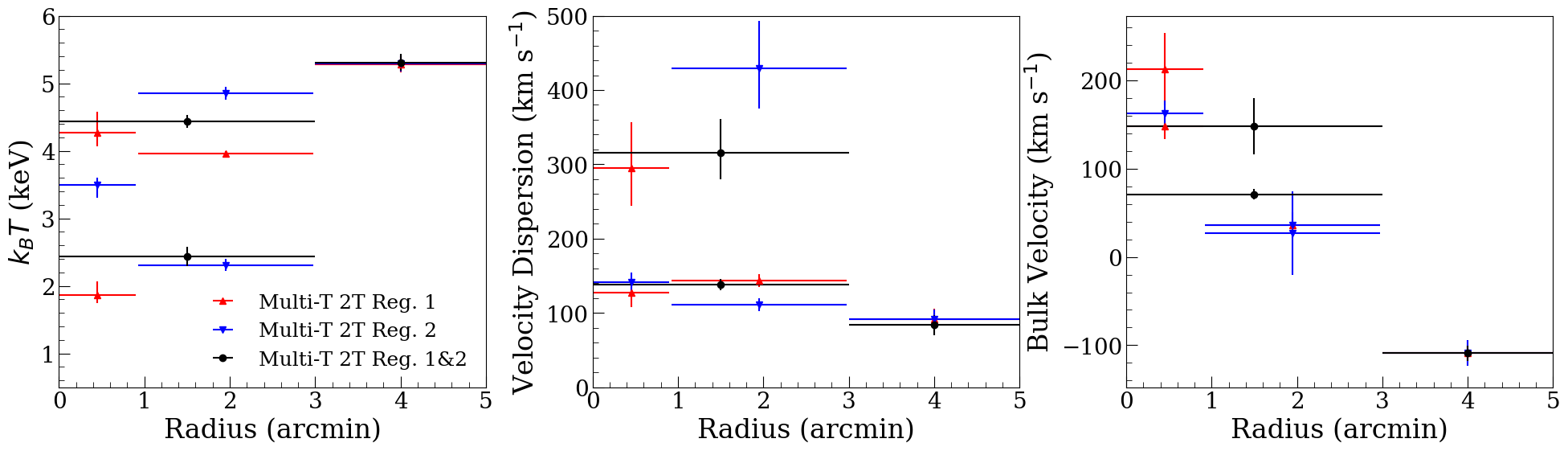}
    \caption{Radial profiles in regions 1--3 of temperature (left), velocity dispersion (middle), and bulk velocity (right) between the three two-temperature models. The red data points represent the values from the model with two temperatures in region 1 (Table \ref{tab:SSM_full_fit_2Treg1}). The blue data points represent the values from the model with two temperatures in region 2 (Table \ref{tab:SSM_full_fit}). The black data points represent the values from the model with two temperatures in the combined region (Table \ref{tab:SSM_combined_fit}). The bulk velocities are calculated as $v_\mathrm{bulk}=c(z-z_\mathrm{BCG})/(1+z_\mathrm{BCG})$, relative to the BCG redshift $z_\mathrm{BCG}=0.017284$, with heliocentric correction. The redshift values, $z$, are from the Xspec best-fit model.}
    \label{fig:multiT_profiles_compared}
\end{figure*}

While the two-temperature model components trace different metals at different temperatures, they also indicate significantly different velocity dispersions. Figure \ref{fig:multiT_profiles} compares radial profiles of the temperature, velocity dispersion, and bulk velocity of the best-fit model using two temperatures in region 2 and the \citetalias{XRISM_Perseus}. The measured velocity dispersions from region 1 and hotter component of region 2 are systematically lower than the \citetalias{XRISM_Perseus}.  The cooler component of region 2 yields a large 400--500 km/s velocity dispersion. The profiles of the multiple temperature models (Tables \ref{tab:SSM_combined_fit}--\ref{tab:SSM_full_fit_2Treg1}) in the inner three regions are compared in Figure \ref{fig:multiT_profiles_compared}. The discrepancy between the single-temperature hotter component and the \citetalias{XRISM_Perseus} in regions 1 and 2 may result from the two components tracing different emission line features. A one-temperature model would measure slightly larger broadening to account for the wider wings, while the two-temperature model is able to separate the velocity dispersions into a narrower peak component and broader wing component.

\subsection{The cooler gas }\label{subsec:cold_comp}
%The cold component is real
The cooler and hotter component's normalizations are comparable. The cooler component is detected in the central $\sim60$ kpc independent of the extraction regions in the spectral analysis. Two temperatures are statistically preferred over a single temperature model up to a significance of $\Delta\mathrm{BIC}=78$, and the broad component is evident in the emission lines. The modeling challenges of constraining two components in regions 1 and 2 simultaneously raise concerns whether the second temperature component might be a modeling artifact, especially considering the complexity of the model. However, it is robustly detected against other systematic effects (see section \ref{subsec:systematics}). While we cannot completely rule out a modeling artifact, the cooler component appears to be physically real and we now explore its properties.

\subsubsection{Dynamics and energetics}

The broad velocity dispersion, cooler gas is detected in region 2 with the highest statistical significance and the highest velocity dispersion.  It therefore provides upper bounds on the dynamics of the cooler component. The sound speed is given by
\begin{equation}
    c_s = \sqrt{\gamma\frac{k_BT}{\mu m_p}},
\end{equation}
where $\gamma=5/3$ is the adiabatic index, $T$ is the temperature, $\mu=0.61$ is the mean molecular weight, and $m_p$ is the proton mass. The sound speed of the cooler $k_BT=2.30^{+0.11}_{-0.08}$ keV gas is $c_s=776^{+19}_{-13}$ km\,s$^{-1}$. The measured one dimensional velocity dispersion, $\sigma=428^{+64}_{-54}$ km\,s$^{-1}$, is close to the sound speed. Assuming the velocity dispersion is due to isotropic turbulence, the 3-D Mach number is 
\begin{equation}
    M_{3\mathrm{D}}=\sqrt{3}\,\sigma/c_s,
\end{equation}
where $\sigma$ is the measured velocity dispersion. Therefore, $M_{3\mathrm{D}}\simeq0.96\pm0.14$ suggests the turbulence is mildly transonic. However, the measured velocity dispersion is likely not purely turbulence as region 2 contains bubbles, sloshing, and other features that can induce peculiar motions \citep[e.g.][]{FabianPerseus2006}. Nevertheless, taken at face value, the cooler component's 3D Mach number is much higher than the hot component ($M_{3\mathrm{D}}\sim$ 0.16). The non-thermal pressure fraction is given by
\begin{equation}
    \frac{P_\mathrm{NT}}{P_\mathrm{tot}} = \frac{M_{3\mathrm{D}}^2}{M_{3\mathrm{D}}^2 + 3/\gamma}.
\end{equation}
The cooler component's non-thermal pressure fraction would be $\sim33.7\pm6.8\%$, which is significantly larger than the hotter component ($\sim 1.6\%$) in Perseus
and in other objects \citep{XRISM_2025_A2029,XRISM_Centaurus_2025,XRISM2025Coma,XRISM_Ophiuchus2025,XRISM2025HydraA}.
The cooling time, the timescale for the gas to radiate away its thermal energy, is given by
\begin{equation}
    t_\mathrm{cool}=\frac{3}{2}\frac{Nk_BT}{n_in_e\Lambda(T)},
\end{equation}
where $N$ is the total number density of all species and $\Lambda(T)$ is the normalized cooling function. We assume $\Lambda(T)=1.86\times10^{-23}$ erg cm$^{3}$ s$^{-1}$ for solar metallicity gas at $k_BT=2.30$ keV \citep{Coolingfunction1993}. We further assume the relative normalizations between the hotter and cooler components represent the relative fractions of the total gas mass. Then, using the $n_e$ profile from \cite{Tang2017}, we calculate $t_\mathrm{cool}\sim1.1$ Gyr for the cooler component.

The age of the inner bubbles located within regions 1 and 2 are estimated to be $10^{7-8}$ yr \citep{Fabian2000}. Assuming the cooler gas was lifted from the center of the galaxy at the speed of the velocity dispersion $\sigma=428^{+64}_{-54}$ km\,s$^{-1}$, which is close to the buoyancy terminal speed, its crossing times, $t_\mathrm{cross} \simeq r/\sigma$, through regions 1 and 2 would be relatively short.  For region 1, which is $\sim20$ kpc in radius $t_\mathrm{cross}\sim4.6\times10^7$ yr.  The crossing time from the cluster center halfway through region 2 is $t_\mathrm{cross}\sim9.1\times10^7$ yr and $t_\mathrm{cross}\sim1.4\times10^8$ yr to transit region 2. Under these assumptions it is plausible for the cooler gas to have been lifted throughout regions  1 and 2 by the bubbles. 

The energy to create the cavities is estimated to be $P\Delta V\simeq8\times10^{58}$ erg \citep{Fabian2000}. Assuming the cooler gas fills a spherical volume of radius $R$ with the velocity dispersion representing the 1D velocity of the gas, we can then estimate the kinetic energy as $\mathrm{K.E.}=\frac{3}{2}m\sigma^2$, where $m=4\pi\int_0^R\rho(r)r^2dr$. The gas mass of the cooler component is found as \citep{Hogan2017}
\begin{equation}\label{eq:n_e}
    n_e=D_A(1+z)10^7\sqrt{\frac{\mathrm{Norm}\times4\,\pi\,1.2}{V}}.
\end{equation}
Calculating $n_e$ from equation \ref{eq:n_e}, and using the $n_e$ profile reported by \cite{Tang2017}, we calculate the range of kinetic energy to be $\simeq4.4-12.3\times10^{58}$ erg for a sphere of 20 kpc radius and $\simeq3.5-7.3\times10^{59}$ erg for a sphere of 40 kpc radius. The assumption that the kinetic energy scales with a factor of $3/2$ may not be accurate as the velocity is likely not gaussian random motion, so these calculations are uncertain by a factor of $\sim2$ or more. However, within these broad uncertainties the energies of the bubbles and the cooler gas component are comparable.

The above calculations assume that the cooler gas is lifted at the velocity dispersion speed. Turbulence is likely generated in situ, so driving the radial outflow of the cooler gas at the velocity dispersion speed through turbulence may not be physical. However, rising bubbles form eddies in their wakes, which entrain the surrounding gas \citep{Churazov2001}. Numerical simulations show the entrained gas rises at speeds comparable to or larger than the bubble-rise velocity \citep{Congyao2022}. The velocity dispersion is close to the buoyancy terminal speed, so the assumption that the cooler gas rises at the velocity dispersion speed is plausible.

\subsubsection{Physical interpretation of the Cooler Gas}
\begin{figure}[ht]
    \centering
    \includegraphics[width=\columnwidth]{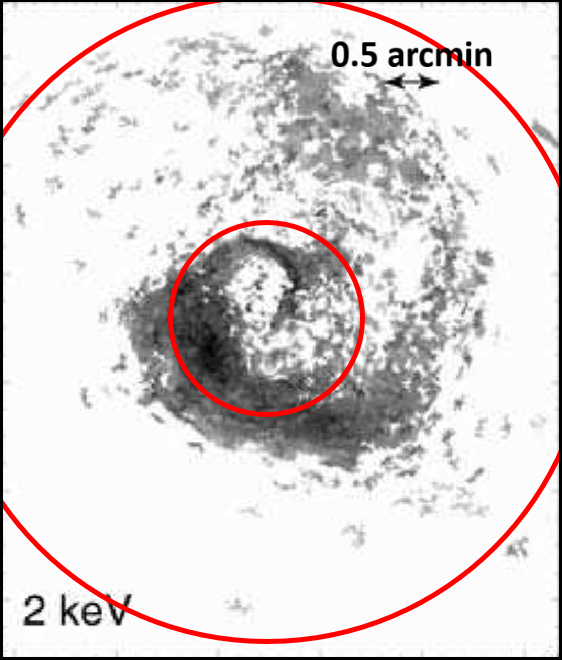}
    \caption{Regions 1 and 2 of the multiple temperature analysis overlaid on the 2 keV Chandra mass map by \citet{FabianPerseus2006}.}
    \label{fig:massmap}
\end{figure}

The origin of the cooler gas is unknown but its existence represents a significant development in cluster atmospheric physics. It may have been introduced by a merger. The 2 keV Chandra mass map  \cite{FabianPerseus2006} shows a spiral structure wrapping around the cluster center with a larger radius of curvature and broader width at higher altitudes. Figure \ref{fig:massmap} shows the map with regions 1 and 2 roughly aligned, showing that this cool spiral lies entirely within these regions. The temperature of the gas in this mass map is close to that of the cooler gas discovered here, and much of the gas lies within the region where we detect it with highest significance.

Large velocity dispersions, comparable to that of the cooler component, are found in merger simulations. These mergers show spiral structure similar to what is seen in Perseus \citep{Zuhone2016}. The simulations predict significant deviations from Gaussianity that can be modeled as a superposition of Gaussian line spread functions, which is how the two-temperature model fits emission lines (e.g. S {\scriptsize XVI} Ly$\alpha$, Fig. \ref{fig:multiT_lines_comparison}). Although the high velocity dispersions in the simulations are associated with gas roughly 2--3 times the temperatures we observe, the cooler sloshing gas in the simulations always has a larger velocity dispersion than the hotter gas, which is qualitatively similar to our findings. 

Alternatively, the AGN may be having a stronger impact on cooler gas than the warmer gas. Since the cooler component is only detected in the inner regions, it may be cooling gas that has been disturbed by the jets and bubbles.
Cool gas has been detected around the rims of the inner bubbles \citep{Fabian2003,FabianPerseus2006}. Rising bubbles drag cooler gas from the core \citep{Churazov2001,Kirkpatrick2011}. Within the broad uncertainties of the analysis, the dynamics and energetics of the cooler gas are consistent with that of the bubbles. Therefore it is plausible that the cooler component is cooling gas that has been disturbed and perhaps lifted outward by the jets and bubbles.

The central galaxy is surrounded by a large nebula and additional filamentary structures of multiphase gas, associated with warm ($\sim10^4$ K) ionized gas (e.g. \citealt{Fabian2008}) that encompasses much of the region where the cooler gas component is detected. 
This gas reports low velocity dispersions of $\lesssim50$ km s$^{-1}$ beyond the inner $\sim20$ kpc \citep{Gendron2018,Rhea2025}. If the cooler hot gas component represents cooling gas being disturbed by the jets and bubbles, then this implies that the $10^4$ K gas and the cooler component ($\sim10^{7.4}$ K) have significantly different kinematics.  Given that their volume filling factors and gas densities differ by several order of magnitude, this may not be surprising. 

Despite detecting the cooler component in regions 1 and 2 using multiple approaches, that the cooler gas is detected with lower significance in region 1 is concerning. However, uncertainties in modeling the bright AGN point source coupled with the broad PSF adds a great deal of complexity to the thermal model of this region. This coupled with insufficient count statistics to nail down a coupled model across all regions yielding a stable solution.
Moreover, the detected cooler component may not be the 2 keV gas shown in Figure \ref{fig:massmap}, and therefore may not be distributed similarly. Temperature maps from deep Chandra observations of Perseus show a more extended distribution of gas at 2 keV throughout the entire central $\sim60$ kpc (see Fig. 19 in \citealt{FabianPerseus2007}). If we assume that the cooler gas is dragged by the bubbles and potentially influenced by the merger dynamics beyond the central region, it is possible that less cool X-ray gas at the detected temperature resides in the core than in region 2, although we give this interpretation less weight.

\section{Conclusion}\label{Conclusion}
We performed an analysis of the five XRISM Resolve observations of the Perseus cluster. We measured emission line flux ratios and performed spectral fitting with multiple temperature models to measure the temperature structure of the Perseus cluster. Flux ratios of Si--Fe are consistent with single temperature models, while multiple temperature spectral fitting indicates significant atmospheric gas components with at least two temperatures in the central pointing within $\sim 60$ kpc. 

The cooler gas component features a high velocity dispersion, potentially indicating mildly transonic turbulence and a high non-thermal pressure fraction. The kinematics of the cooler gas component are significantly different from the one-temperature and hotter component counterparts. The cool gas may be associated with a  sloshing spiral of gas generated by a recent merger, and/or the AGN preferentially impacting the cooler gas.
However, the potentially complex kinematics and structure expected from both physical mechanisms coupled with the relatively primitive information available for the cooler component do not permit a clearer interpretation.  The combined effects of the sloshing spiral and the central jet activity may be at work.

Deeper XRISM observations of the central Perseus Cluster targeting higher signal at softer energies will be necessary to reach clearer conclusions. XRISM results are rapidly emerging. Detections of cool, broad velocity-width gas have been reported in the cluster atmospheres of Cygnus A \citep{XRISM_CygnusA} and M87 \citep{Simionescu_2026}.  If ubiquitous and resilient to systematic errors associated with the gain calibration, cool, broad gas components may be an important new element in our understanding of cooling flows, feedback, and cluster weather.

%% Also note that the akcnowlodgment environment does not support long amounts of text. If you have a lot of people and institutions to acknowledge, do not use this command. Instead, create a new \section{Acknowledgments}.
\begin{acknowledgments}
% CSA & NSERC acknowledgement
BRM and JM acknowledge the Canadian Space Agency and the Natural Sciences and Engineering Research Council of Canada for generous support.
% Congyao's acknowledgement
CZ acknowledges the support of the Czech Science Foundation (GACR) Junior Star grant no. GM24-10599M.
% XRISM acknowledgement
This work was supported by JSPS Core-to-Core Program, (grant number: JPJSCCA20220002).
% HEASARC acknowledgement
This research has made use of data and/or software provided by the High Energy Astrophysics Science Archive Research Center (HEASARC), which is a service of the Astrophysics Science Division at NASA/GSFC.
% Chandra data acknowledgement
This research has made use of data obtained from the Chandra Data Archive provided by the Chandra X-ray Center (CXC).
% NASA acknowledgement
The material is based upon work supported by NASA under award number 80GSFC24M0006.
\end{acknowledgments}


\vspace{5mm}
\facilities{XRISM(Resolve), Chandra(ACIS)}


\software{HEASoft,
          Xspec \& PyXspec \citep{Xspec},
          Astropy \citep{astropy},
          Matplotlib \citep{matplotlib},
          NumPy \citep{numpy_2011,harris_array_2020},
          Python \citep{van_rossum_python_2009},
          CIAO \citep{CIAO}
          }



\appendix

\section{Line Ratios Single-Temperature Baseline Models}\label{appendix:LineRatios}
Provided in Table \ref{tab:lineratios_base} below are the baseline \texttt{bapec} models used in the narrow band fits for the line ratio analysis for the C0 and C1 pointings (section \ref{subsec:Line Ratios}). The models are obtained by fitting the full pointing spectrum over 1.8--10 keV.

\begin{table}[h]
    \centering
    \caption{Single-temperature baseline models for the line ratio narrow band fits for the C0 and C1 pointings.}
    \renewcommand{\arraystretch}{1.25}
    \begin{tabular}{lcc}
    \hline
    Parameter & C0 & C1 \\
    \hline
    $k_BT$ (keV) & $3.67\pm0.03$ & $4.44\pm0.08$\\
    Abundance & $0.63\pm0.01$ & $0.54\pm0.02$\\
    z ($\times10^{-2}$) & $1.767\pm0.001$ & $1.721\pm0.002$\\
    $\sigma$ (km\,s$^{-1}$) & $173\pm3$ & $136\pm6$\\
    \hline
    \end{tabular}
    \label{tab:lineratios_base}
\end{table}

\section{Alternative AGN Fit}\label{appendix:AGN}
Provided in Table \ref{tab:SSM_full_fit_appendix} below is the full profile fit as described in section \ref{subsec:multi-T}, but where the power law photon index is frozen instead of the flux.

\begin{table*}[h]
    \centering
    \caption{Best fit parameters of all sub-FOV regions with SSM from 1.8--10 keV, with two temperatures in region 2, with power law $\Gamma$ frozen instead of flux. Details of each region model are provided in sections \ref{subsec:multi-T} and \ref{subsec:systematics:agn}. Some abundances were not constrainable in the outer regions, so those parameters were frozen at solar values. The power law flux best-fit value is $\mathrm{Flux}_{2-10\mathrm{\,keV}}=42^{+1}_{-1}\times10^{-12}$ erg cm$^{-2}$ s$^{-1}$. Fit C-stat/dof = 156503/163923.}
    \renewcommand{\arraystretch}{1.25}
    \begin{tabular*}{\linewidth}{@{\extracolsep{\fill}} lccccc}
    \hline
    Parameter & Region 1 & Region 2 & Region 3 & Region 4 & Region 5 \\
    \hline
    Radius (arcmin) & 0 -- 0.9 & 0.9 -- 3 & 3 -- 5 & 5 -- 7 & 7 -- 11.85 \\
    
    $k_BT$ (keV) & $2.93^{+0.09}_{-0.09}$ & $4.82^{+0.10}_{-0.10}$ & $5.27^{+0.12}_{-0.12}$ & $5.89^{+0.18}_{-0.17}$ & $7.42^{+0.26}_{-0.26}$\\
    
    Fe abund. & $0.88^{+0.05}_{-0.04}$ & $0.74^{+0.02}_{-0.02}$ & $0.48^{+0.02}_{-0.02}$ & $0.54^{+0.03}_{-0.03}$ & $0.36^{+0.03}_{-0.02}$\\
    
    z ($\times10^{-2}$) & $1.792^{+0.003}_{-0.003}$ & $1.752^{+0.002}_{-0.002}$ & $1.700^{+0.003}_{-0.003}$ & $1.701^{+0.005}_{-0.005}$ & $1.777^{+0.008}_{-0.008}$\\
    
    $\sigma$ (km\,s$^{-1}$) & $134^{+13}_{-13}$ & $111^{+8}_{-9}$ & $90^{+13}_{-14}$ & $177^{+18}_{-17}$ & $202^{+25}_{-23}$\\
    
    Norm ($\times10^{-2}$) & $6.51^{+0.20}_{-0.21}$ & $16.15^{+0.81}_{-0.88}$ & $17.87^{+0.35}_{-0.35}$ & $3.87^{+0.09}_{-0.09}$ & $3.84^{+0.07}_{-0.07}$\\
    
    ${k_BT}_2$ (keV) & -- & $2.30^{+0.11}_{-0.08}$ & -- & -- & --\\
    
    z$_2$ ($\times10^{-2}$) & -- & $1.732^{+0.019}_{-0.020}$ & -- & -- & --\\
    
    $\sigma_2$ (km\,s$^{-1}$) & -- & $464^{+75}_{-65}$ & -- & -- & --\\
    
    Norm$_2$ ($\times10^{-2}$) & -- & $12.38^{+1.04}_{-1.09}$ & -- & -- & --\\
    
    \hline
    \end{tabular*}
    \label{tab:SSM_full_fit_appendix}
\end{table*}


\bibliography{sample631}{}

@INPROCEEDINGS{Xspec,
       author = {{Arnaud}, K.~A.},
        title = "{XSPEC: The First Ten Years}",
    booktitle = {Astronomical Data Analysis Software and Systems V},
         year = 1996,
       editor = {{Jacoby}, George H. and {Barnes}, Jeannette},
       series = {Astronomical Society of the Pacific Conference Series},
       volume = {101},
        month = jan,
        pages = {17},
       adsurl = {https://ui.adsabs.harvard.edu/abs/1996ASPC..101...17A}
}

@ARTICLE{Cash,
       author = {{Cash}, W.},
        title = "{Parameter estimation in astronomy through application of the likelihood ratio.}",
      journal = {\apj},
         year = 1979,
        month = mar,
       volume = {228},
        pages = {939-947},
          doi = {10.1086/156922},
       adsurl = {https://ui.adsabs.harvard.edu/abs/1979ApJ...228..939C}
}

@ARTICLE{tbabs,
       author = {{Wilms}, J. and {Allen}, A. and {McCray}, R.},
        title = "{On the Absorption of X-Rays in the Interstellar Medium}",
      journal = {\apj},
         year = 2000,
        month = oct,
       volume = {542},
       number = {2},
        pages = {914-924},
          doi = {10.1086/317016},
archivePrefix = {arXiv},
       eprint = {astro-ph/0008425},
 primaryClass = {astro-ph},
       adsurl = {https://ui.adsabs.harvard.edu/abs/2000ApJ...542..914W}
}

@article{atomdb,
doi = {10.1088/0004-637X/756/2/128},
url = {https://dx.doi.org/10.1088/0004-637X/756/2/128},
year = {2012},
month = {aug},
publisher = {The American Astronomical Society},
volume = {756},
number = {2},
pages = {128},
author = {Foster, A. R. and Ji, L. and Smith, R. K. and Brickhouse, N. S.},
title = {UPDATED ATOMIC DATA AND CALCULATIONS FOR X-RAY SPECTROSCOPY},
journal = {The Astrophysical Journal}
}

@ARTICLE{Lodders2009,
       author = {{Lodders}, K. and {Palme}, H. and {Gail}, H. -P.},
        title = "{Abundances of the Elements in the Solar System}",
      journal = {Landolt B{\"o}rnstein},
         year = 2009,
        month = jan,
       volume = {4B},
        pages = {712},
          doi = {10.1007/978-3-540-88055-4_34},
archivePrefix = {arXiv},
       eprint = {0901.1149},
 primaryClass = {astro-ph.EP},
       adsurl = {https://ui.adsabs.harvard.edu/abs/2009LanB...4B..712L}
}

@article{Perseus_Hitomi_2018,
    author = {{Hitomi Collaboration} and Aharonian, Felix and Akamatsu, Hiroki and Akimoto, Fumie and Allen, Steven W and Angelini, Lorella and Audard, Marc and Awaki, Hisamitsu and Axelsson, Magnus and Bamba, Aya and Bautz, Marshall W and Blandford, Roger and Brenneman, Laura W and Brown, Gregory V and Bulbul, Esra and Cackett, Edward M and Canning, Rebecca E A and Chernyakova, Maria and Chiao, Meng P and Coppi, Paolo S and Costantini, Elisa and de Plaa, Jelle and de Vries, Cor P and den Herder, Jan-Willem and Done, Chris and Dotani, Tadayasu and Ebisawa, Ken and Eckart, Megan E and Enoto, Teruaki and Ezoe, Yuichiro and Fabian, Andrew C and Ferrigno, Carlo and Foster, Adam R and Fujimoto, Ryuichi and Fukazawa, Yasushi and Furuzawa, Akihiro and Galeazzi, Massimiliano and Gallo, Luigi C and Gandhi, Poshak and Giustini, Margherita and Goldwurm, Andrea and Gu, Liyi and Guainazzi, Matteo and Haba, Yoshito and Hagino, Kouichi and Hamaguchi, Kenji and Harrus, Ilana M and Hatsukade, Isamu and Hayashi, Katsuhiro and Hayashi, Takayuki and Hayashi, Tasuku and Hayashida, Kiyoshi and Hiraga, Junko S and Hornschemeier, Ann and Hoshino, Akio and Hughes, John P and Ichinohe, Yuto and Iizuka, Ryo and Inoue, Hajime and Inoue, Shota and Inoue, Yoshiyuki and Ishida, Manabu and Ishikawa, Kumi and Ishisaki, Yoshitaka and Iwai, Masachika and Kaastra, Jelle and Kallman, Tim and Kamae, Tsuneyoshi and Kataoka, Jun and Katsuda, Satoru and Kawai, Nobuyuki and Kelley, Richard L and Kilbourne, Caroline A and Kitaguchi, Takao and Kitamoto, Shunji and Kitayama, Tetsu and Kohmura, Takayoshi and Kokubun, Motohide and Koyama, Katsuji and Koyama, Shu and Kretschmar, Peter and Krimm, Hans A and Kubota, Aya and Kunieda, Hideyo and Laurent, Philippe and Lee, Shiu-Hang and Leutenegger, Maurice A and Limousin, Olivier and Loewenstein, Michael and Long, Knox S and Lumb, David and Madejski, Greg and Maeda, Yoshitomo and Maier, Daniel and Makishima, Kazuo and Markevitch, Maxim and Matsumoto, Hironori and Matsushita, Kyoko and McCammon, Dan and McNamara, Brian R and Mehdipour, Missagh and Miller, Eric D and Miller, Jon M and Mineshige, Shin and Mitsuda, Kazuhisa and Mitsuishi, Ikuyuki and Miyazawa, Takuya and Mizuno, Tsunefumi and Mori, Hideyuki and Mori, Koji and Mukai, Koji and Murakami, Hiroshi and Mushotzky, Richard F and Nakagawa, Takao and Nakajima, Hiroshi and Nakamori, Takeshi and Nakashima, Shinya and Nakazawa, Kazuhiro and Nobukawa, Kumiko K and Nobukawa, Masayoshi and Noda, Hirofumi and Odaka, Hirokazu and Ohashi, Takaya and Ohno, Masanori and Okajima, Takashi and Ota, Naomi and Ozaki, Masanobu and Paerels, Frits and Paltani, Stéphane and Petre, Robert and Pinto, Ciro and Porter, Frederick S and Pottschmidt, Katja and Reynolds, Christopher S and Safi-Harb, Samar and Saito, Shinya and Sakai, Kazuhiro and Sasaki, Toru and Sato, Goro and Sato, Kosuke and Sato, Rie and Sawada, Makoto and Schartel, Norbert and Serlemtsos, Peter J and Seta, Hiromi and Shidatsu, Megumi and Simionescu, Aurora and Smith, Randall K and Soong, Yang and Stawarz, Łukasz and Sugawara, Yasuharu and Sugita, Satoshi and Szymkowiak, Andrew and Tajima, Hiroyasu and Takahashi, Hiromitsu and Takahashi, Tadayuki and Takeda, Shin’ichiro and Takei, Yoh and Tamagawa, Toru and Tamura, Takayuki and Tanaka, Keigo and Tanaka, Takaaki and Tanaka, Yasuo and Tanaka, Yasuyuki T and Tashiro, Makoto S and Tawara, Yuzuru and Terada, Yukikatsu and Terashima, Yuichi and Tombesi, Francesco and Tomida, Hiroshi and Tsuboi, Yohko and Tsujimoto, Masahiro and Tsunemi, Hiroshi and Tsuru, Takeshi Go and Uchida, Hiroyuki and Uchiyama, Hideki and Uchiyama, Yasunobu and Ueda, Shutaro and Ueda, Yoshihiro and Uno, Shin’ichiro and Urry, C Megan and Ursino, Eugenio and Wang, Qian H S and Watanabe, Shin and Werner, Norbert and Wilkins, Dan R and Williams, Brian J and Yamada, Shinya and Yamaguchi, Hiroya and Yamaoka, Kazutaka and Yamasaki, Noriko Y and Yamauchi, Makoto and Yamauchi, Shigeo and Yaqoob, Tahir and Yatsu, Yoichi and Yonetoku, Daisuke and Zhuravleva, Irina and Zoghbi, Abderahmen},
    title = {Atmospheric gas dynamics in the Perseus cluster observed with Hitomi*},
    journal = {Publications of the Astronomical Society of Japan},
    volume = {70},
    number = {2},
    pages = {9},
    year = {2018},
    month = {04},
    issn = {0004-6264},
    doi = {10.1093/pasj/psx138},
    url = {https://doi.org/10.1093/pasj/psx138},
    eprint = {https://academic.oup.com/pasj/article-pdf/70/2/9/54668816/pasj\_70\_2\_9.pdf},
}

@article{Hitomi_Perseus_lineratios,
    author = {{Hitomi Collaboration} and Aharonian, Felix and Akamatsu, Hiroki and Akimoto, Fumie and Allen, Steven W and Angelini, Lorella and Audard, Marc and Awaki, Hisamitsu and Axelsson, Magnus and Bamba, Aya and Bautz, Marshall W and Blandford, Roger and Brenneman, Laura W and Brown, Gregory V and Bulbul, Esra and Cackett, Edward M and Chernyakova, Maria and Chiao, Meng P and Coppi, Paolo S and Costantini, Elisa and de Plaa, Jelle and de Vries, Cor P and den Herder, Jan-Willem and Done, Chris and Dotani, Tadayasu and Ebisawa, Ken and Eckart, Megan E and Enoto, Teruaki and Ezoe, Yuichiro and Fabian, Andrew C and Ferrigno, Carlo and Foster, Adam R and Fujimoto, Ryuichi and Fukazawa, Yasushi and Furukawa, Maki and Furuzawa, Akihiro and Galeazzi, Massimiliano and Gallo, Luigi C and Gandhi, Poshak and Giustini, Margherita and Goldwurm, Andrea and Gu, Liyi and Guainazzi, Matteo and Haba, Yoshito and Hagino, Kouichi and Hamaguchi, Kenji and Harrus, Ilana M and Hatsukade, Isamu and Hayashi, Katsuhiro and Hayashi, Takayuki and Hayashida, Kiyoshi and Hiraga, Junko S and Hornschemeier, Ann and Hoshino, Akio and Hughes, John P and Ichinohe, Yuto and Iizuka, Ryo and Inoue, Hajime and Inoue, Yoshiyuki and Ishida, Manabu and Ishikawa, Kumi and Ishisaki, Yoshitaka and Iwai, Masachika and Kaastra, Jelle and Kallman, Tim and Kamae, Tsuneyoshi and Kataoka, Jun and Kato, Yuichi and Katsuda, Satoru and Kawai, Nobuyuki and Kelley, Richard L and Kilbourne, Caroline A and Kitaguchi, Takao and Kitamoto, Shunji and Kitayama, Tetsu and Kohmura, Takayoshi and Kokubun, Motohide and Koyama, Katsuji and Koyama, Shu and Kretschmar, Peter and Krimm, Hans A and Kubota, Aya and Kunieda, Hideyo and Laurent, Philippe and Lee, Shiu-Hang and Leutenegger, Maurice A and Limousin, Olivier and Loewenstein, Michael and Long, Knox S and Lumb, David and Madejski, Greg and Maeda, Yoshitomo and Maier, Daniel and Makishima, Kazuo and Markevitch, Maxim and Matsumoto, Hironori and Matsushita, Kyoko and McCammon, Dan and McNamara, Brian R and Mehdipour, Missagh and Miller, Eric D and Miller, Jon M and Mineshige, Shin and Mitsuda, Kazuhisa and Mitsuishi, Ikuyuki and Miyazawa, Takuya and Mizuno, Tsunefumi and Mori, Hideyuki and Mori, Koji and Mukai, Koji and Murakami, Hiroshi and Mushotzky, Richard F and Nakagawa, Takao and Nakajima, Hiroshi and Nakamori, Takeshi and Nakashima, Shinya and Nakazawa, Kazuhiro and Nobukawa, Kumiko K and Nobukawa, Masayoshi and Noda, Hirofumi and Odaka, Hirokazu and Ohashi, Takaya and Ohno, Masanori and Okajima, Takashi and Ota, Naomi and Ozaki, Masanobu and Paerels, Frits and Paltani, Stéphane and Petre, Robert and Pinto, Ciro and Porter, Frederick S and Pottschmidt, Katja and Reynolds, Christopher S and Safi-Harb, Samar and Saito, Shinya and Sakai, Kazuhiro and Sasaki, Toru and Sato, Goro and Sato, Kosuke and Sato, Rie and Sawada, Makoto and Schartel, Norbert and Serlemtsos, Peter J and Seta, Hiromi and Shidatsu, Megumi and Simionescu, Aurora and Smith, Randall K and Soong, Yang and Stawarz, Łukasz and Sugawara, Yasuharu and Sugita, Satoshi and Szymkowiak, Andrew and Tajima, Hiroyasu and Takahashi, Hiromitsu and Takahashi, Tadayuki and Takeda, Shiníchiro and Takei, Yoh and Tamagawa, Toru and Tamura, Takayuki and Tanaka, Takaaki and Tanaka, Yasuo and Tanaka, Yasuyuki T and Tashiro, Makoto S and Tawara, Yuzuru and Terada, Yukikatsu and Terashima, Yuichi and Tombesi, Francesco and Tomida, Hiroshi and Tsuboi, Yohko and Tsujimoto, Masahiro and Tsunemi, Hiroshi and Tsuru, Takeshi Go and Uchida, Hiroyuki and Uchiyama, Hideki and Uchiyama, Yasunobu and Ueda, Shutaro and Ueda, Yoshihiro and Uno, Shiníchiro and Urry, C Megan and Ursino, Eugenio and Watanabe, Shin and Werner, Norbert and Wilkins, Dan R and Williams, Brian J and Yamada, Shinya and Yamaguchi, Hiroya and Yamaoka, Kazutaka and Yamasaki, Noriko Y and Yamauchi, Makoto and Yamauchi, Shigeo and Yaqoob, Tahir and Yatsu, Yoichi and Yonetoku, Daisuke and Zhuravleva, Irina and Zoghbi, Abderahmen},
    title = {Temperature structure in the Perseus cluster core observed with Hitomi*},
    journal = {Publications of the Astronomical Society of Japan},
    volume = {70},
    number = {2},
    pages = {11},
    year = {2018},
    month = {04},
    issn = {0004-6264},
    doi = {10.1093/pasj/psy004},
    url = {https://doi.org/10.1093/pasj/psy004},
    eprint = {https://academic.oup.com/pasj/article-pdf/70/2/11/54669070/pasj\_70\_2\_11.pdf},
}

@article{XRISM_flightdata2025,
author = {Frederick Scott Porter and Caroline A. Kilbourne and Meng P. Chiao and Renata S. Cumbee and Megan E. Eckart and Ryuichi Fujimoto and Yoshitaka Ishisaki and Yoshiaki Kanemaru and Richard L. Kelley and Maurice Andrew Leutenegger and Yoshitomo Maeda and Misaki Mizumoto and Kosuke Sato and Makoto Sawada and Gary A. Sneiderman and Yoh Takei and Masahiro Tsujimoto and Yuusuke Uchida and Tomomi Watanabe and Shin’ya Yamada},
title = {{In-flight performance of the XRISM/Resolve detector system}},
volume = {11},
journal = {Journal of Astronomical Telescopes, Instruments, and Systems},
number = {4},
publisher = {SPIE},
pages = {042016},
year = {2025},
doi = {10.1117/1.JATIS.11.4.042016},
URL = {https://doi.org/10.1117/1.JATIS.11.4.042016}
}

@ARTICLE{Perseus_Hitomi_2016,
       author = {{Hitomi Collaboration} and {Aharonian}, Felix and {Akamatsu}, Hiroki and {Akimoto}, Fumie and {Allen}, Steven W. and {Anabuki}, Naohisa and {Angelini}, Lorella and {Arnaud}, Keith and {Audard}, Marc and {Awaki}, Hisamitsu and {Axelsson}, Magnus and {Bamba}, Aya and {Bautz}, Marshall and {Blandford}, Roger and {Brenneman}, Laura and {Brown}, Gregory V. and {Bulbul}, Esra and {Cackett}, Edward and {Chernyakova}, Maria and {Chiao}, Meng and {Coppi}, Paolo and {Costantini}, Elisa and {de Plaa}, Jelle and {den Herder}, Jan-Willem and {Done}, Chris and {Dotani}, Tadayasu and {Ebisawa}, Ken and {Eckart}, Megan and {Enoto}, Teruaki and {Ezoe}, Yuichiro and {Fabian}, Andrew C. and {Ferrigno}, Carlo and {Foster}, Adam and {Fujimoto}, Ryuichi and {Fukazawa}, Yasushi and {Furuzawa}, Akihiro and {Galeazzi}, Massimiliano and {Gallo}, Luigi and {Gandhi}, Poshak and {Giustini}, Margherita and {Goldwurm}, Andrea and {Gu}, Liyi and {Guainazzi}, Matteo and {Haba}, Yoshito and {Hagino}, Kouichi and {Hamaguchi}, Kenji and {Harrus}, Ilana and {Hatsukade}, Isamu and {Hayashi}, Katsuhiro and {Hayashi}, Takayuki and {Hayashida}, Kiyoshi and {Hiraga}, Junko and {Hornschemeier}, Ann and {Hoshino}, Akio and {Hughes}, John and {Iizuka}, Ryo and {Inoue}, Hajime and {Inoue}, Yoshiyuki and {Ishibashi}, Kazunori and {Ishida}, Manabu and {Ishikawa}, Kumi and {Ishisaki}, Yoshitaka and {Itoh}, Masayuki and {Iyomoto}, Naoko and {Kaastra}, Jelle and {Kallman}, Timothy and {Kamae}, Tuneyoshi and {Kara}, Erin and {Kataoka}, Jun and {Katsuda}, Satoru and {Katsuta}, Junichiro and {Kawaharada}, Madoka and {Kawai}, Nobuyuki and {Kelley}, Richard and {Khangulyan}, Dmitry and {Kilbourne}, Caroline and {King}, Ashley and {Kitaguchi}, Takao and {Kitamoto}, Shunji and {Kitayama}, Tetsu and {Kohmura}, Takayoshi and {Kokubun}, Motohide and {Koyama}, Shu and {Koyama}, Katsuji and {Kretschmar}, Peter and {Krimm}, Hans and {Kubota}, Aya and {Kunieda}, Hideyo and {Laurent}, Philippe and {Lebrun}, Fran{\c{c}}ois and {Lee}, Shiu-Hang and {Leutenegger}, Maurice and {Limousin}, Olivier and {Loewenstein}, Michael and {Long}, Knox S. and {Lumb}, David and {Madejski}, Grzegorz and {Maeda}, Yoshitomo and {Maier}, Daniel and {Makishima}, Kazuo and {Markevitch}, Maxim and {Matsumoto}, Hironori and {Matsushita}, Kyoko and {McCammon}, Dan and {McNamara}, Brian and {Mehdipour}, Missagh and {Miller}, Eric and {Miller}, Jon and {Mineshige}, Shin and {Mitsuda}, Kazuhisa and {Mitsuishi}, Ikuyuki and {Miyazawa}, Takuya and {Mizuno}, Tsunefumi and {Mori}, Hideyuki and {Mori}, Koji and {Moseley}, Harvey and {Mukai}, Koji and {Murakami}, Hiroshi and {Murakami}, Toshio and {Mushotzky}, Richard and {Nagino}, Ryo and {Nakagawa}, Takao and {Nakajima}, Hiroshi and {Nakamori}, Takeshi and {Nakano}, Toshio and {Nakashima}, Shinya and {Nakazawa}, Kazuhiro and {Nobukawa}, Masayoshi and {Noda}, Hirofumi and {Nomachi}, Masaharu and {O'Dell}, Steve and {Odaka}, Hirokazu and {Ohashi}, Takaya and {Ohno}, Masanori and {Okajima}, Takashi and {Ota}, Naomi and {Ozaki}, Masanobu and {Paerels}, Frits and {Paltani}, Stephane and {Parmar}, Arvind and {Petre}, Robert and {Pinto}, Ciro and {Pohl}, Martin and {Porter}, F. Scott and {Pottschmidt}, Katja and {Ramsey}, Brian and {Reynolds}, Christopher and {Russell}, Helen and {Safi-Harb}, Samar and {Saito}, Shinya and {Sakai}, Kazuhiro and {Sameshima}, Hiroaki and {Sato}, Goro and {Sato}, Kosuke and {Sato}, Rie and {Sawada}, Makoto and {Schartel}, Norbert and {Serlemitsos}, Peter and {Seta}, Hiromi and {Shidatsu}, Megumi and {Simionescu}, Aurora and {Smith}, Randall and {Soong}, Yang and {Stawarz}, Lukasz and {Sugawara}, Yasuharu and {Sugita}, Satoshi and {Szymkowiak}, Andrew and {Tajima}, Hiroyasu and {Takahashi}, Hiromitsu and {Takahashi}, Tadayuki and {Takeda}, Shin'Ichiro and {Takei}, Yoh and {Tamagawa}, Toru and {Tamura}, Keisuke and {Tamura}, Takayuki and {Tanaka}, Takaaki and {Tanaka}, Yasuo and {Tanaka}, Yasuyuki and {Tashiro}, Makoto and {Tawara}, Yuzuru and {Terada}, Yukikatsu and {Terashima}, Yuichi and {Tombesi}, Francesco and {Tomida}, Hiroshi and {Tsuboi}, Yohko and {Tsujimoto}, Masahiro and {Tsunemi}, Hiroshi and {Tsuru}, Takeshi and {Uchida}, Hiroyuki and {Uchiyama}, Hideki and {Uchiyama}, Yasunobu and {Ueda}, Shutaro and {Ueda}, Yoshihiro and {Ueno}, Shiro and {Uno}, Shin'Ichiro and {Urry}, Meg and {Ursino}, Eugenio and {de Vries}, Cor and {Watanabe}, Shin and {Werner}, Norbert and {Wik}, Daniel and {Wilkins}, Dan and {Williams}, Brian and {Yamada}, Shinya and {Yamaguchi}, Hiroya and {Yamaoka}, Kazutaka and {Yamasaki}, Noriko Y. and {Yamauchi}, Makoto and {Yamauchi}, Shigeo and {Yaqoob}, Tahir and {Yatsu}, Yoichi and {Yonetoku}, Daisuke and {Yoshida}, Atsumasa and {Yuasa}, Takayuki and {Zhuravleva}, Irina and {Zoghbi}, Abderahmen},
        title = "{The quiescent intracluster medium in the core of the Perseus cluster}",
      journal = {Nature},
         year = {2016},
        month = {jul},
       volume = {535},
       number = {7610},
        pages = {117-121},
          doi = {10.1038/nature18627},
archivePrefix = {arXiv},
       eprint = {1607.04487},
 primaryClass = {astro-ph.GA},
       adsurl = {https://ui.adsabs.harvard.edu/abs/2016Natur.535..117H}
}

@ARTICLE{FabianPerseus2007,
       author = {{Sanders}, J.~S. and {Fabian}, A.~C.},
        title = "{A deeper X-ray study of the core of the Perseus galaxy cluster: the power of sound waves and the distribution of metals and cosmic rays}",
      journal = {\mnras},
         year = 2007,
        month = nov,
       volume = {381},
       number = {4},
        pages = {1381-1399},
          doi = {10.1111/j.1365-2966.2007.12347.x},
archivePrefix = {arXiv},
       eprint = {0705.2712},
 primaryClass = {astro-ph},
       adsurl = {https://ui.adsabs.harvard.edu/abs/2007MNRAS.381.1381S}
}

@ARTICLE{XRISM_Ophiuchus2025,
       author = {{Fujita}, Yutaka and {Fukushima}, Kotaro and {Sato}, Kosuke and {Fukazawa}, Yasushi and {Kondo}, Marie},
        title = "{XRISM Observation of the Ophiuchus Galaxy Cluster: Quiescent Velocity Structure in the Dynamically Disturbed Core}",
      journal = {arXiv e-prints},
         year = 2025,
        month = jun,
          eid = {arXiv:2507.00126},
        pages = {arXiv:2507.00126},
          doi = {10.48550/arXiv.2507.00126},
archivePrefix = {arXiv},
       eprint = {2507.00126},
 primaryClass = {astro-ph.HE},
       adsurl = {https://ui.adsabs.harvard.edu/abs/2025arXiv250700126F}
}

@ARTICLE{XRISM_Centaurus_2025,
       author = {{XRISM Collaboration} and {Audard}, Marc and {Awaki}, Hisamitsu and {Ballhausen}, Ralf and {Bamba}, Aya and {Behar}, Ehud and {Boissay-Malaquin}, Rozenn and {Brenneman}, Laura and {Brown}, Gregory V. and {Corrales}, Lia and {Costantini}, Elisa and {Cumbee}, Renata and {Done}, Chris and {Dotani}, Tadayasu and {Ebisawa}, Ken and {Eckart}, Megan E. and {Eckert}, Dominique and {Enoto}, Teruaki and {Eguchi}, Satoshi and {Ezoe}, Yuichiro and {Foster}, Adam and {Fujimoto}, Ryuichi and {Fujita}, Yutaka and {Fukazawa}, Yasushi and {Fukushima}, Kotaro and {Furuzawa}, Akihiro and {Gallo}, Luigi and {Garc{\'\i}a}, Javier A. and {Gu}, Liyi and {Guainazzi}, Matteo and {Hagino}, Kouichi and {Hamaguchi}, Kenji and {Hatsukade}, Isamu and {Hayashi}, Katsuhiro and {Hayashi}, Takayuki and {Hell}, Natalie and {Hodges-Kluck}, Edmund and {Hornschemeier}, Ann and {Ichinohe}, Yuto and {Ishida}, Manabu and {Ishikawa}, Kumi and {Ishisaki}, Yoshitaka and {Kaastra}, Jelle and {Kallman}, Timothy and {Kara}, Erin and {Katsuda}, Satoru and {Kanemaru}, Yoshiaki and {Kelley}, Richard and {Kilbourne}, Caroline and {Kitamoto}, Shunji and {Kobayashi}, Shogo and {Kohmura}, Takayoshi and {Kubota}, Aya and {Leutenegger}, Maurice and {Loewenstein}, Michael and {Maeda}, Yoshitomo and {Markevitch}, Maxim and {Matsumoto}, Hironori and {Matsushita}, Kyoko and {McCammon}, Dan and {McNamara}, Brian and {Mernier}, Fran{\c{c}}ois and {Miller}, Eric D. and {Miller}, Jon M. and {Mitsuishi}, Ikuyuki and {Mizumoto}, Misaki and {Mizuno}, Tsunefumi and {Mori}, Koji and {Mukai}, Koji and {Murakami}, Hiroshi and {Mushotzky}, Richard and {Nakajima}, Hiroshi and {Nakazawa}, Kazuhiro and {Ness}, Jan-Uwe and {Nobukawa}, Kumiko and {Nobukawa}, Masayoshi and {Noda}, Hirofumi and {Odaka}, Hirokazu and {Ogawa}, Shoji and {Ogorzalek}, Anna and {Okajima}, Takashi and {Ota}, Naomi and {Paltani}, Stephane and {Petre}, Robert and {Plucinsky}, Paul and {Porter}, Frederick Scott and {Pottschmidt}, Katja and {Sato}, Kosuke and {Sato}, Toshiki and {Sawada}, Makoto and {Seta}, Hiromi and {Shidatsu}, Megumi and {Simionescu}, Aurora and {Smith}, Randall and {Suzuki}, Hiromasa and {Szymkowiak}, Andrew and {Takahashi}, Hiromitsu and {Takeo}, Mai and {Tamagawa}, Toru and {Tamura}, Keisuke and {Tanaka}, Takaaki and {Tanimoto}, Atsushi and {Tashiro}, Makoto and {Terada}, Yukikatsu and {Terashima}, Yuichi and {Trigo}, Mar{\'\i}a D{\'\i}az and {Tsuboi}, Yohko and {Tsujimoto}, Masahiro and {Tsunemi}, Hiroshi and {Tsuru}, Takeshi G. and {Uchida}, Hiroyuki and {Uchida}, Nagomi and {Uchida}, Yuusuke and {Uchiyama}, Hideki and {Ueda}, Yoshihiro and {Uno}, Shinichiro and {Vink}, Jacco and {Watanabe}, Shin and {Williams}, Brian J. and {Yamada}, Satoshi and {Yamada}, Shinya and {Yamaguchi}, Hiroya and {Yamaoka}, Kazutaka and {Yamasaki}, Noriko Y. and {Yamauchi}, Makoto and {Yamauchi}, Shigeo and {Yaqoob}, Tahir and {Yoneyama}, Tomokage and {Yoshida}, Tessei and {Yukita}, Mihoko and {Zhuravleva}, Irina and {Kondo}, Marie and {Werner}, Norbert and {Pl{\v{s}}ek}, Tom{\'a}{\v{s}} and {Sun}, Ming and {Hosogi}, Kokoro and {Majumder}, Anwesh},
        title = "{The bulk motion of gas in the core of the Centaurus galaxy cluster}",
      journal = {\nat},
         year = 2025,
        month = feb,
       volume = {638},
       number = {8050},
        pages = {365-369},
          doi = {10.1038/s41586-024-08561-z},
archivePrefix = {arXiv},
       eprint = {2502.08722},
 primaryClass = {astro-ph.HE},
       adsurl = {https://ui.adsabs.harvard.edu/abs/2025Natur.638..365X}
}

@article{XRISM_2025_A2029,
       author = {{XRISM Collaboration} and {Audard}, Marc and {Awaki}, Hisamitsu and {Ballhausen}, Ralf and {Bamba}, Aya and {Behar}, Ehud and {Boissay-Malaquin}, Rozenn and {Brenneman}, Laura and {Brown}, Gregory V. and {Corrales}, Lia and {Costantini}, Elisa and {Cumbee}, Renata and {Diaz Trigo}, Maria and {Done}, Chris and {Dotani}, Tadayasu and {Ebisawa}, Ken and {Eckart}, Megan E. and {Eckert}, Dominique and {Eguchi}, Satoshi and {Enoto}, Teruaki and {Ezoe}, Yuichiro and {Foster}, Adam and {Fujimoto}, Ryuichi and {Fujita}, Yutaka and {Fukazawa}, Yasushi and {Fukushima}, Kotaro and {Furuzawa}, Akihiro and {Gallo}, Luigi and {Garc{\'\i}a}, Javier A. and {Gu}, Liyi and {Guainazzi}, Matteo and {Hagino}, Kouichi and {Hamaguchi}, Kenji and {Hatsukade}, Isamu and {Hayashi}, Katsuhiro and {Hayashi}, Takayuki and {Hell}, Natalie and {Hodges-Kluck}, Edmund and {Hornschemeier}, Ann and {Ichinohe}, Yuto and {Ishida}, Manabu and {Ishikawa}, Kumi and {Ishisaki}, Yoshitaka and {Kaastra}, Jelle and {Kallman}, Timothy and {Kara}, Erin and {Katsuda}, Satoru and {Kanemaru}, Yoshiaki and {Kelley}, Richard and {Kilbourne}, Caroline and {Kitamoto}, Shunji and {Kobayashi}, Shogo and {Kohmura}, Takayoshi and {Kubota}, Aya and {Leutenegger}, Maurice and {Loewenstein}, Michael and {Maeda}, Yoshitomo and {Markevitch}, Maxim and {Matsumoto}, Hironori and {Matsushita}, Kyoko and {McCammon}, Dan and {McNamara}, Brian and {Mernier}, Fran{\c{c}}ois and {Miller}, Eric D. and {Miller}, Jon M. and {Mitsuishi}, Ikuyuki and {Mizumoto}, Misaki and {Mizuno}, Tsunefumi and {Mori}, Koji and {Mukai}, Koji and {Murakami}, Hiroshi and {Mushotzky}, Richard and {Nakajima}, Hiroshi and {Nakazawa}, Kazuhiro and {Ness}, Jan-Uwe and {Nobukawa}, Kumiko and {Nobukawa}, Masayoshi and {Noda}, Hirofumi and {Odaka}, Hirokazu and {Ogawa}, Shoji and {Ogorzalek}, Anna and {Okajima}, Takashi and {Ota}, Naomi and {Paltani}, Stephane and {Petre}, Robert and {Plucinsky}, Paul and {Porter}, Frederick S. and {Pottschmidt}, Katja and {Sato}, Kosuke and {Sato}, Toshiki and {Sawada}, Makoto and {Seta}, Hiromi and {Shidatsu}, Megumi and {Simionescu}, Aurora and {Smith}, Randall and {Suzuki}, Hiromasa and {Szymkowiak}, Andrew and {Takahashi}, Hiromitsu and {Takeo}, Mai and {Tamagawa}, Toru and {Tamura}, Keisuke and {Tanaka}, Takaaki and {Tanimoto}, Atsushi and {Tashiro}, Makoto and {Terada}, Yukikatsu and {Terashima}, Yuichi and {Tsuboi}, Yohko and {Tsujimoto}, Masahiro and {Tsunemi}, Hiroshi and {Tsuru}, Takeshi and {Uchida}, Hiroyuki and {Uchida}, Nagomi and {Uchida}, Yuusuke and {Uchiyama}, Hideki and {Ueda}, Yoshihiro and {Uno}, Shinichiro and {Vink}, Jacco and {Watanabe}, Shin and {Williams}, Brian J. and {Yamada}, Satoshi and {Yamada}, Shinya and {Yamaguchi}, Hiroya and {Yamaoka}, Kazutaka and {Yamasaki}, Noriko and {Yamauchi}, Makoto and {Yamauchi}, Shigeo and {Yaqoob}, Tahir and {Yoneyama}, Tomokage and {Yoshida}, Tessei and {Yukita}, Mihoko and {Zhuravleva}, Irina and {Bartalesi}, Tommaso and {Ettori}, Stefano and {Kosarzycki}, Roman and {Lovisari}, Lorenzo and {Rose}, Tom and {Sarkar}, Arnab and {Sun}, Ming and {Tamhane}, Prathamesh},
        title = "{XRISM Reveals Low Nonthermal Pressure in the Core of the Hot, Relaxed Galaxy Cluster A2029}",
      journal = {\apjl},
         year = 2025,
        month = mar,
       volume = {982},
       number = {1},
          eid = {L5},
        pages = {L5},
          doi = {10.3847/2041-8213/ada7cd},
archivePrefix = {arXiv},
       eprint = {2501.05514},
 primaryClass = {astro-ph.HE},
       adsurl = {https://ui.adsabs.harvard.edu/abs/2025ApJ...982L...5X}
}

@ARTICLE{XRISM2025Coma,
       author = {{XRISM Collaboration} and {Audard}, Marc and {Awaki}, Hisamitsu and {Ballhausen}, Ralf and {Bamba}, Aya and {Behar}, Ehud and {Boissay-Malaquin}, Rozenn and {Brenneman}, Laura and {Brown}, Gregory V. and {Corrales}, Lia and {Costantini}, Elisa and {Cumbee}, Renata and {Diaz Trigo}, Maria and {Done}, Chris and {Dotani}, Tadayasu and {Ebisawa}, Ken and {Eckart}, Megan E. and {Eckert}, Dominique and {Eguchi}, Satoshi and {Enoto}, Teruaki and {Ezoe}, Yuichiro and {Foster}, Adam and {Fujimoto}, Ryuichi and {Fujita}, Yutaka and {Fukazawa}, Yasushi and {Fukushima}, Kotaro and {Furuzawa}, Akihiro and {Gallo}, Luigi and {Garcia}, Javier A. and {Gu}, Liyi and {Guainazzi}, Matteo and {Hagino}, Kouichi and {Hamaguchi}, Kenji and {Hatsukade}, Isamu and {Hayashi}, Katsuhiro and {Hayashi}, Takayuki and {Hell}, Natalie and {Hodges-Kluck}, Edmund and {Hornschemeier}, Ann and {Ichinohe}, Yuto and {Ishi}, Daiki and {Ishida}, Manabu and {Ishikawa}, Kumi and {Ishisaki}, Yoshitaka and {Kaastra}, Jelle and {Kallman}, Timothy and {Kara}, Erin and {Katsuda}, Satoru and {Kanemaru}, Yoshiaki and {Kelley}, Richard and {Kilbourne}, Caroline and {Kitamoto}, Shunji and {Kobayashi}, Shogo and {Kohmura}, Takayoshi and {Kubota}, Aya and {Leutenegger}, Maurice and {Loewenstein}, Michael and {Maeda}, Yoshitomo and {Markevitch}, Maxim and {Matsumoto}, Hironori and {Matsushita}, Kyoko and {McCammon}, Dan and {McNamara}, Brian and {Mernier}, Francois and {Miller}, Eric D. and {Miller}, Jon M. and {Mitsuishi}, Ikuyuki and {Mizumoto}, Misaki and {Mizuno}, Tsunefumi and {Mori}, Koji and {Mukai}, Koji and {Murakami}, Hiroshi and {Mushotzky}, Richard and {Nakajima}, Hiroshi and {Nakazawa}, Kazuhiro and {Ness}, Jan-Uwe and {Nobukawa}, Kumiko and {Nobukawa}, Masayoshi and {Noda}, Hirofumi and {Odaka}, Hirokazu and {Ogawa}, Shoji and {Ogorzalek}, Anna and {Okajima}, Takashi and {Ota}, Naomi and {Paltani}, Stephane and {Petre}, Robert and {Plucinsky}, Paul and {Porter}, Frederick S. and {Pottschmidt}, Katja and {Sato}, Kosuke and {Sato}, Toshiki and {Sawada}, Makoto and {Seta}, Hiromi and {Shidatsu}, Megumi and {Simionescu}, Aurora and {Smith}, Randall and {Suzuki}, Hiromasa and {Szymkowiak}, Andrew and {Takahashi}, Hiromitsu and {Takeo}, Mai and {Tamagawa}, Toru and {Tamura}, Keisuke and {Tanaka}, Takaaki and {Tanimoto}, Atsushi and {Tashiro}, Makoto and {Terada}, Yukikatsu and {Terashima}, Yuichi and {Tsuboi}, Yohko and {Tsujimoto}, Masahiro and {Tsunemi}, Hiroshi and {Tsuru}, Takeshi and {Tumer}, Aysegul and {Uchida}, Hiroyuki and {Uchida}, Nagomi and {Uchida}, Yuusuke and {Uchiyama}, Hideki and {Ueda}, Shutaro and {Ueda}, Yoshihiro and {Uno}, Shinichiro and {Vink}, Jacco and {Watanabe}, Shin and {Williams}, Brian J. and {Yamada}, Satoshi and {Yamada}, Shinya and {Yamaguchi}, Hiroya and {Yamaoka}, Kazutaka and {Yamasaki}, Noriko and {Yamauchi}, Makoto and {Yamauchi}, Shigeo and {Yaqoob}, Tahir and {Yoneyama}, Tomokage and {Yoshida}, Tessei and {Yukita}, Mihoko and {Zhuravleva}, Irina and {Fabian}, Andrew and {Nelson}, Dylan and {Okabe}, Nobuhiro and {Pillepich}, Annalisa and {Potter}, Cicely and {Regamey}, Manon and {Sakai}, Kosei and {Shishido}, Mona and {Truong}, Nhut and {Wik}, Daniel R. and {ZuHone}, John},
        title = "{XRISM forecast for the Coma cluster: stormy, with a steep power spectrum}",
      journal = {arXiv e-prints},
         year = 2025,
        month = apr,
          eid = {arXiv:2504.20928},
        pages = {arXiv:2504.20928},
archivePrefix = {arXiv},
       eprint = {2504.20928},
 primaryClass = {astro-ph.HE},
       adsurl = {https://ui.adsabs.harvard.edu/abs/2025arXiv250420928X}
}

@ARTICLE{XRISM2025HydraA,
       author = {{Rose}, Tom and {McNamara}, B.~R. and {Meunier}, Julian and {Fabian}, A.~C. and {Russell}, Helen and {Nulsen}, Paul and {Dizdar}, Neo and {Heckman}, Timothy M. and {McDonald}, Michael and {Markevitch}, Maxim and {Paerels}, Frits and {Simionescu}, Aurora and {Werner}, Norbert and {Coil}, Alison L. and {Hodges-Kluck}, Edmund and {Miller}, Eric D. and {Wise}, Michael},
        title = "{A XRISM Observation of the Archetypal Radio-Mode Feedback System Hydra-A: Measurements of Atmospheric Motion and Constraints on Turbulent Dissipation}",
      journal = {arXiv e-prints},
         year = 2025,
        month = may,
          eid = {arXiv:2505.01494},
        pages = {arXiv:2505.01494},
          doi = {10.48550/arXiv.2505.01494},
archivePrefix = {arXiv},
       eprint = {2505.01494},
 primaryClass = {astro-ph.GA},
       adsurl = {https://ui.adsabs.harvard.edu/abs/2025arXiv250501494R}
}

@article{Churazov2003,
doi = {10.1086/374923},
url = {https://dx.doi.org/10.1086/374923},
year = {2003},
month = {jun},
publisher = {},
volume = {590},
number = {1},
pages = {225},
author = {Churazov, E. and Forman, W. and Jones, C. and Böhringer, H.},
title = {XMM-Newton Observations of the Perseus Cluster. I. The Temperature and Surface Brightness Structure},
journal = {The Astrophysical Journal}
}

@ARTICLE{Coolingfunction1993,
       author = {{Sutherland}, Ralph S. and {Dopita}, M.~A.},
        title = "{Cooling Functions for Low-Density Astrophysical Plasmas}",
      journal = {\apjs},
         year = 1993,
        month = sep,
       volume = {88},
        pages = {253},
          doi = {10.1086/191823},
       adsurl = {https://ui.adsabs.harvard.edu/abs/1993ApJS...88..253S}
}

@ARTICLE{FabianPerseus2006,
       author = {{Fabian}, A.~C. and {Sanders}, J.~S. and {Taylor}, G.~B. and {Allen}, S.~W. and {Crawford}, C.~S. and {Johnstone}, R.~M. and {Iwasawa}, K.},
        title = "{A very deep Chandra observation of the Perseus cluster: shocks, ripples and conduction}",
      journal = {\mnras},
         year = 2006,
        month = feb,
       volume = {366},
       number = {2},
        pages = {417-428},
          doi = {10.1111/j.1365-2966.2005.09896.x},
archivePrefix = {arXiv},
       eprint = {astro-ph/0510476},
 primaryClass = {astro-ph},
       adsurl = {https://ui.adsabs.harvard.edu/abs/2006MNRAS.366..417F}
}

@article{astropy,
	author = {{The Astropy Collaboration} and {Robitaille, Thomas P.} and {Tollerud, Erik J.} and {Greenfield, Perry} and {Droettboom, Michael} and {Bray, Erik} and {Aldcroft, Tom} and {Davis, Matt} and {Ginsburg, Adam} and {Price-Whelan, Adrian M.} and {Kerzendorf, Wolfgang E.} and {Conley, Alexander} and {Crighton, Neil} and {Barbary, Kyle} and {Muna, Demitri} and {Ferguson, Henry} and {Grollier, Frédéric} and {Parikh, Madhura M.} and {Nair, Prasanth H.} and {Günther, Hans M.} and {Deil, Christoph} and {Woillez, Julien} and {Conseil, Simon} and {Kramer, Roban} and {Turner, James E. H.} and {Singer, Leo} and {Fox, Ryan} and {Weaver, Benjamin A.} and {Zabalza, Victor} and {Edwards, Zachary I.} and {Azalee Bostroem, K.} and {Burke, D. J.} and {Casey, Andrew R.} and {Crawford, Steven M.} and {Dencheva, Nadia} and {Ely, Justin} and {Jenness, Tim} and {Labrie, Kathleen} and {Lim, Pey Lian} and {Pierfederici, Francesco} and {Pontzen, Andrew} and {Ptak, Andy} and {Refsdal, Brian} and {Servillat, Mathieu} and {Streicher, Ole}},
	title = {Astropy: A community Python package for astronomy},
	DOI= "10.1051/0004-6361/201322068",
	url= "https://doi.org/10.1051/0004-6361/201322068",
	journal = {A\&A},
	year = 2013,
	volume = 558,
	pages = "A33",
	month = "",
}

@ARTICLE{matplotlib,
  author={Hunter, John D.},
  journal={Computing in Science \& Engineering}, 
  title={Matplotlib: A 2D Graphics Environment}, 
  year={2007},
  volume={9},
  number={3},
  pages={90-95},
  doi={10.1109/MCSE.2007.55}}

@book{van_rossum_python_2009,
  title = {Python 3 {{Reference Manual}}},
  author = {Van Rossum, Guido and Drake, Fred L.},
  year = {2009},
  publisher = {{CreateSpace}},
  address = {{Scotts Valley, CA}},
  isbn = {978-1-4414-1269-0}
}

@article{numpy_2011,
  title = {The {{NumPy Array}}: {{A Structure}} for {{Efficient Numerical Computation}}},
  shorttitle = {The {{NumPy Array}}},
  author = {van der Walt, S. and Colbert, S. C. and Varoquaux, G.},
  year = {2011},
  month = mar,
  volume = {13},
  pages = {22--30},
  issn = {1521-9615},
  doi = {10.1109/MCSE.2011.37},
  journal = {Comput. Sci. Eng.},
  number = {2}
}

@article{harris_array_2020,
  title = {Array Programming with {{NumPy}}},
  author = {Harris, Charles R. and Millman, K. Jarrod and {van der Walt}, St{\'e}fan J. and Gommers, Ralf and Virtanen, Pauli and Cournapeau, David and Wieser, Eric and Taylor, Julian and Berg, Sebastian and Smith, Nathaniel J. and Kern, Robert and Picus, Matti and Hoyer, Stephan and {van Kerkwijk}, Marten H. and Brett, Matthew and Haldane, Allan and {del R{\'i}o}, Jaime Fern{\'a}ndez and Wiebe, Mark and Peterson, Pearu and {G{\'e}rard-Marchant}, Pierre and Sheppard, Kevin and Reddy, Tyler and Weckesser, Warren and Abbasi, Hameer and Gohlke, Christoph and Oliphant, Travis E.},
  year = {2020},
  month = sep,
  volume = {585},
  pages = {357--362},
  publisher = {{Nature Publishing Group}},
  issn = {1476-4687},
  doi = {10.1038/s41586-020-2649-2},
  url = {https://www.nature.com/articles/s41586-020-2649-2},
  urldate = {2020-09-17},
  journal = {Nature},
  language = {en},
  number = {7825}
}

@article{Perseus_Hitomi_ResonantScattering,
    author = {{Hitomi Collaboration} and Aharonian, Felix and Akamatsu, Hiroki and Akimoto, Fumie and Allen, Steven W and Angelini, Lorella and Audard, Marc and Awaki, Hisamitsu and Axelsson, Magnus and Bamba, Aya and Bautz, Marshall W and Blandford, Roger and Brenneman, Laura W and Brown, Gregory V and Bulbul, Esra and Cackett, Edward M and Chernyakova, Maria and Chiao, Meng P and Coppi, Paolo S and Costantini, Elisa and de Plaa, Jelle and de Vries, Cor P and den Herder, Jan-Willem and Done, Chris and Dotani, Tadayasu and Ebisawa, Ken and Eckart, Megan E and Enoto, Teruaki and Ezoe, Yuichiro and Fabian, Andrew C and Ferrigno, Carlo and Foster, Adam R and Fujimoto, Ryuichi and Fukazawa, Yasushi and Furukawa, Maki and Furuzawa, Akihiro and Galeazzi, Massimiliano and Gallo, Luigi C and Gandhi, Poshak and Giustini, Margherita and Goldwurm, Andrea and Gu, Liyi and Guainazzi, Matteo and Haba, Yoshito and Hagino, Kouichi and Hamaguchi, Kenji and Harrus, Ilana M and Hatsukade, Isamu and Hayashi, Katsuhiro and Hayashi, Takayuki and Hayashida, Kiyoshi and Hiraga, Junko S and Hornschemeier, Ann and Hoshino, Akio and Hughes, John P and Ichinohe, Yuto and Iizuka, Ryo and Inoue, Hajime and Inoue, Yoshiyuki and Ishida, Manabu and Ishikawa, Kumi and Ishisaki, Yoshitaka and Iwai, Masachika and Kaastra, Jelle and Kallman, Tim and Kamae, Tsuneyoshi and Kataoka, Jun and Katsuda, Satoru and Kawai, Nobuyuki and Kelley, Richard L and Kilbourne, Caroline A and Kitaguchi, Takao and Kitamoto, Shunji and Kitayama, Tetsu and Kohmura, Takayoshi and Kokubun, Motohide and Koyama, Katsuji and Koyama, Shu and Kretschmar, Peter and Krimm, Hans A and Kubota, Aya and Kunieda, Hideyo and Laurent, Philippe and Lee, Shiu-Hang and Leutenegger, Maurice A and Limousin, Olivier O and Loewenstein, Michael and Long, Knox S and Lumb, David and Madejski, Greg and Maeda, Yoshitomo and Maier, Daniel and Makishima, Kazuo and Markevitch, Maxim and Matsumoto, Hironori and Matsushita, Kyoko and McCammon, Dan and McNamara, Brian R and Mehdipour, Missagh and Miller, Eric D and Miller, Jon M and Mineshige, Shin and Mitsuda, Kazuhisa and Mitsuishi, Ikuyuki and Miyazawa, Takuya and Mizuno, Tsunefumi and Mori, Hideyuki and Mori, Koji and Mukai, Koji and Murakami, Hiroshi and Mushotzky, Richard F and Nakagawa, Takao and Nakajima, Hiroshi and Nakamori, Takeshi and Nakashima, Shinya and Nakazawa, Kazuhiro and Nobukawa, Kumiko K and Nobukawa, Masayoshi and Noda, Hirofumi and Odaka, Hirokazu and Ogorzalek, Anna and Ohashi, Takaya and Ohno, Masanori and Okajima, Takashi and Ota, Naomi and Ozaki, Masanobu and Paerels, Frits and Paltani, Stéphane and Petre, Robert and Pinto, Ciro and Porter, Frederick S and Pottschmidt, Katja and Reynolds, Christopher S and Safi-Harb, Samar and Saito, Shinya and Sakai, Kazuhiro and Sasaki, Toru and Sato, Goro and Sato, Kosuke and Sato, Rie and Sawada, Makoto and Schartel, Norbert and Serlemtsos, Peter J and Seta, Hiromi and Shidatsu, Megumi and Simionescu, Aurora and Smith, Randall K and Soong, Yang and Stawarz, Łukasz and Sugawara, Yasuharu and Sugita, Satoshi and Szymkowiak, Andrew and Tajima, Hiroyasu and Takahashi, Hiromitsu and Takahashi, Tadayuki and Takeda, Shiníchiro and Takei, Yoh and Tamagawa, Toru and Tamura, Takayuki and Tanaka, Takaaki and Tanaka, Yasuo and Tanaka, Yasuyuki T and Tashiro, Makoto S and Tawara, Yuzuru and Terada, Yukikatsu and Terashima, Yuichi and Tombesi, Francesco and Tomida, Hiroshi and Tsuboi, Yohko and Tsujimoto, Masahiro and Tsunemi, Hiroshi and Tsuru, Takeshi Go and Uchida, Hiroyuki and Uchiyama, Hideki and Uchiyama, Yasunobu and Ueda, Shutaro and Ueda, Yoshihiro and Uno, Shiníchiro and Urry, C Megan and Ursino, Eugenio and Watanabe, Shin and Werner, Norbert and Wilkins, Dan R and Williams, Brian J and Yamada, Shinya and Yamaguchi, Hiroya and Yamaoka, Kazutaka and Yamasaki, Noriko Y and Yamauchi, Makoto and Yamauchi, Shigeo and Yaqoob, Tahir and Yatsu, Yoichi and Yonetoku, Daisuke and Zhuravleva, Irina and Zoghbi, Abderahmen},
    title = {Measurements of resonant scattering in the Perseus Cluster core with Hitomi SXS*},
    journal = {Publications of the Astronomical Society of Japan},
    volume = {70},
    number = {2},
    pages = {10},
    year = {2018},
    month = {04},
    issn = {0004-6264},
    doi = {10.1093/pasj/psx127},
    url = {https://doi.org/10.1093/pasj/psx127},
    eprint = {https://academic.oup.com/pasj/article-pdf/70/2/10/54668509/pasj\_70\_2\_10.pdf},
}

@INPROCEEDINGS{CIAO,
       author = {{Fruscione}, Antonella and {McDowell}, Jonathan C. and {Allen}, Glenn E. and {Brickhouse}, Nancy S. and {Burke}, Douglas J. and {Davis}, John E. and {Durham}, Nick and {Elvis}, Martin and {Galle}, Elizabeth C. and {Harris}, Daniel E. and {Huenemoerder}, David P. and {Houck}, John C. and {Ishibashi}, Bish and {Karovska}, Margarita and {Nicastro}, Fabrizio and {Noble}, Michael S. and {Nowak}, Michael A. and {Primini}, Frank A. and {Siemiginowska}, Aneta and {Smith}, Randall K. and {Wise}, Michael},
        title = "{CIAO: Chandra's data analysis system}",
    booktitle = {Observatory Operations: Strategies, Processes, and Systems},
         year = 2006,
       editor = {{Silva}, David R. and {Doxsey}, Rodger E.},
       series = {Society of Photo-Optical Instrumentation Engineers (SPIE) Conference Series},
       volume = {6270},
        month = jun,
          eid = {62701V},
        pages = {62701V},
          doi = {10.1117/12.671760},
       adsurl = {https://ui.adsabs.harvard.edu/abs/2006SPIE.6270E..1VF}
}

@ARTICLE{XRISM_CygnusA,
       author = {{Majumder}, Anwesh and {Heckman}, T. and {Meunier}, J. and {Simionescu}, A. and {McNamara}, B.~R. and {Gu}, L. and {Ptak}, A. and {Hodges-Kluck}, E. and {Yukita}, M. and {Wise}, M.~W. and {Roy}, N.},
        title = "{Spectrally Resolved Gas Kinematics in Cygnus A: XRISM Detects AGN Jet-induced Velocity Dispersion in Multi-temperature Gas}",
      journal = {arXiv e-prints},
         year = 2025,
        month = dec,
          eid = {arXiv:2512.10167},
        pages = {arXiv:2512.10167},
          doi = {10.48550/arXiv.2512.10167},
archivePrefix = {arXiv},
       eprint = {2512.10167},
 primaryClass = {astro-ph.HE},
       adsurl = {https://ui.adsabs.harvard.edu/abs/2025arXiv251210167M}
}

@ARTICLE{XRISM_Perseus,
       author = {{XRISM Collaboration} and {Audard}, Marc and {Awaki}, Hisamitsu and {Ballhausen}, Ralf and {Bamba}, Aya and {Behar}, Ehud and {Boissay-Malaquin}, Rozenn and {Brenneman}, Laura and {Brown}, Gregory V. and {Corrales}, Lia and {Costantini}, Elisa and {Cumbee}, Renata and {Diaz Trigo}, Maria and {Done}, Chris and {Dotani}, Tadayasu and {Ebisawa}, Ken and {Eckart}, Megan E. and {Eckert}, Dominique and {Eguchi}, Satoshi and {Enoto}, Teruaki and {Ezoe}, Yuichiro and {Foster}, Adam and {Fujimoto}, Ryuichi and {Fujita}, Yutaka and {Fukazawa}, Yasushi and {Fukushima}, Kotaro and {Furuzawa}, Akihiro and {Gallo}, Luigi and {Garcia}, Javier A. and {Gu}, Liyi and {Guainazzi}, Matteo and {Hagino}, Kouichi and {Hamaguchi}, Kenji and {Hatsukade}, Isamu and {Hayashi}, Katsuhiro and {Hayashi}, Takayuki and {Hell}, Natalie and {Hodges-Kluck}, Edmund and {Hornschemeier}, Ann and {Ichinohe}, Yuto and {Ishi}, Daiki and {Ishida}, Manabu and {Ishikawa}, Kumi and {Ishisaki}, Yoshitaka and {Kaastra}, Jelle and {Kallman}, Timothy and {Kara}, Erin and {Katsuda}, Satoru and {Kanemaru}, Yoshiaki and {Kelley}, Richard and {Kilbourne}, Caroline and {Kitamoto}, Shunji and {Kobayashi}, Shogo and {Kohmura}, Takayoshi and {Kubota}, Aya and {Leutenegger}, Maurice and {Loewenstein}, Michael and {Maeda}, Yoshitomo and {Markevitch}, Maxim and {Matsumoto}, Hironori and {Matsushita}, Kyoko and {McCammon}, Dan and {McNamara}, Brian and {Mernier}, Francois and {Miller}, Eric D. and {Miller}, Jon M. and {Mitsuishi}, Ikuyuki and {Mizumoto}, Misaki and {Mizuno}, Tsunefumi and {Mori}, Koji and {Mukai}, Koji and {Murakami}, Hiroshi and {Mushotzky}, Richard and {Nakajima}, Hiroshi and {Nakazawa}, Kazuhiro and {Ness}, Jan-Uwe and {Nobukawa}, Kumiko and {Nobukawa}, Masayoshi and {Noda}, Hirofumi and {Odaka}, Hirokazu and {Ogawa}, Shoji and {Ogorzalek}, Anna and {Okajima}, Takashi and {Ota}, Naomi and {Paltani}, Stephane and {Petre}, Robert and {Plucinsky}, Paul and {Porter}, Frederick S. and {Pottschmidt}, Katja and {Sato}, Kosuke and {Sato}, Toshiki and {Sawada}, Makoto and {Seta}, Hiromi and {Shidatsu}, Megumi and {Simionescu}, Aurora and {Smith}, Randall and {Suzuki}, Hiromasa and {Szymkowiak}, Andrew and {Takahashi}, Hiromitsu and {Takeo}, Mai and {Tamagawa}, Toru and {Tamura}, Keisuke and {Tanaka}, Takaaki and {Tanimoto}, Atsushi and {Tashiro}, Makoto and {Terada}, Yukikatsu and {Terashima}, Yuichi and {Tsuboi}, Yohko and {Tsujimoto}, Masahiro and {Tsunemi}, Hiroshi and {Tsuru}, Takeshi G. and {Tumer}, Aysegul and {Uchida}, Hiroyuki and {Uchida}, Nagomi and {Uchida}, Yuusuke and {Uchiyama}, Hideki and {Ueda}, Yoshihiro and {Uno}, Shinichiro and {Vink}, Jacco and {Watanabe}, Shin and {Williams}, Brian J. and {Yamada}, Satoshi and {Yamada}, Shinya and {Yamaguchi}, Hiroya and {Yamaoka}, Kazutaka and {Yamasaki}, Noriko and {Yamauchi}, Makoto and {Yamauchi}, Shigeo and {Yaqoob}, Tahir and {Yoneyama}, Tomokage and {Yoshida}, Tessei and {Yukita}, Mihoko and {Zhuravleva}, Irina and {Bellomi}, Elena and {Drury}, Ian and {Heinrich}, Annie and {Hlavacek-Larrondo}, Julie and {Meunier}, Julian and {Migkas}, Kostas and {Shefler}, Lior and {Stancil}, Phillip C. and {Truong}, Nhut and {Ueda}, Shutaro and {Vigneron}, Benjamin and {Zhang}, Congyao and {ZuHone}, John},
        title = "{Disentangling Multiple Gas Kinematic Drivers in the Perseus Galaxy Cluster}",
      journal = {arXiv e-prints},
         year = 2025,
        month = sep,
          eid = {arXiv:2509.04421},
        pages = {arXiv:2509.04421},
          doi = {10.48550/arXiv.2509.04421},
archivePrefix = {arXiv},
       eprint = {2509.04421},
 primaryClass = {astro-ph.HE},
       adsurl = {https://ui.adsabs.harvard.edu/abs/2025arXiv250904421X}
}

@ARTICLE{McNamaraNulsen2007,
       author = {{McNamara}, B.~R. and {Nulsen}, P.~E.~J.},
        title = "{Heating Hot Atmospheres with Active Galactic Nuclei}",
      journal = {\araa},
         year = 2007,
        month = sep,
       volume = {45},
       number = {1},
        pages = {117-175},
          doi = {10.1146/annurev.astro.45.051806.110625},
archivePrefix = {arXiv},
       eprint = {0709.2152},
 primaryClass = {astro-ph},
       adsurl = {https://ui.adsabs.harvard.edu/abs/2007ARA&A..45..117M}
}

@ARTICLE{McNamaraNulsen2012,
       author = {{McNamara}, B.~R. and {Nulsen}, P.~E.~J.},
        title = "{Mechanical feedback from active galactic nuclei in galaxies, groups and clusters}",
      journal = {New Journal of Physics},
         year = 2012,
        month = may,
       volume = {14},
       number = {5},
          eid = {055023},
        pages = {055023},
          doi = {10.1088/1367-2630/14/5/055023},
archivePrefix = {arXiv},
       eprint = {1204.0006},
 primaryClass = {astro-ph.CO},
       adsurl = {https://ui.adsabs.harvard.edu/abs/2012NJPh...14e5023M}
}

@ARTICLE{Fabian2011,
       author = {{Fabian}, A.~C. and {Sanders}, J.~S. and {Allen}, S.~W. and {Canning}, R.~E.~A. and {Churazov}, E. and {Crawford}, C.~S. and {Forman}, W. and {Gabany}, J. and {Hlavacek-Larrondo}, J. and {Johnstone}, R.~M. and {Russell}, H.~R. and {Reynolds}, C.~S. and {Salom{\'e}}, P. and {Taylor}, G.~B. and {Young}, A.~J.},
        title = "{A wide Chandra view of the core of the Perseus cluster}",
      journal = {\mnras},
         year = 2011,
        month = dec,
       volume = {418},
       number = {4},
        pages = {2154-2164},
          doi = {10.1111/j.1365-2966.2011.19402.x},
archivePrefix = {arXiv},
       eprint = {1105.5025},
 primaryClass = {astro-ph.CO},
       adsurl = {https://ui.adsabs.harvard.edu/abs/2011MNRAS.418.2154F}
}

@ARTICLE{Allen2001,
       author = {{Allen}, S.~W. and {Ettori}, S. and {Fabian}, A.~C.},
        title = "{Chandra measurements of the distribution of mass in the luminous lensing cluster Abell 2390}",
      journal = {\mnras},
         year = 2001,
        month = jul,
       volume = {324},
       number = {4},
        pages = {877-890},
          doi = {10.1046/j.1365-8711.2001.04318.x},
archivePrefix = {arXiv},
       eprint = {astro-ph/0008517},
 primaryClass = {astro-ph},
       adsurl = {https://ui.adsabs.harvard.edu/abs/2001MNRAS.324..877A}
}

@ARTICLE{Voit2005,
       author = {{Voit}, G. Mark and {Donahue}, Megan},
        title = "{An Observationally Motivated Framework for AGN Heating of Cluster Cores}",
      journal = {\apj},
         year = 2005,
        month = dec,
       volume = {634},
       number = {2},
        pages = {955-963},
          doi = {10.1086/497063},
archivePrefix = {arXiv},
       eprint = {astro-ph/0509176},
 primaryClass = {astro-ph},
       adsurl = {https://ui.adsabs.harvard.edu/abs/2005ApJ...634..955V}
}

@ARTICLE{Fabian2003,
       author = {{Fabian}, A.~C. and {Sanders}, J.~S. and {Allen}, S.~W. and {Crawford}, C.~S. and {Iwasawa}, K. and {Johnstone}, R.~M. and {Schmidt}, R.~W. and {Taylor}, G.~B.},
        title = "{A deep Chandra observation of the Perseus cluster: shocks and ripples}",
      journal = {\mnras},
         year = 2003,
        month = sep,
       volume = {344},
       number = {3},
        pages = {L43-L47},
          doi = {10.1046/j.1365-8711.2003.06902.x},
archivePrefix = {arXiv},
       eprint = {astro-ph/0306036},
 primaryClass = {astro-ph},
       adsurl = {https://ui.adsabs.harvard.edu/abs/2003MNRAS.344L..43F}
}

@ARTICLE{Zhuravleva2014,
       author = {{Zhuravleva}, I. and {Churazov}, E. and {Schekochihin}, A.~A. and {Allen}, S.~W. and {Ar{\'e}valo}, P. and {Fabian}, A.~C. and {Forman}, W.~R. and {Sanders}, J.~S. and {Simionescu}, A. and {Sunyaev}, R. and {Vikhlinin}, A. and {Werner}, N.},
        title = "{Turbulent heating in galaxy clusters brightest in X-rays}",
      journal = {\nat},
         year = 2014,
        month = nov,
       volume = {515},
       number = {7525},
        pages = {85-87},
          doi = {10.1038/nature13830},
archivePrefix = {arXiv},
       eprint = {1410.6485},
 primaryClass = {astro-ph.HE},
       adsurl = {https://ui.adsabs.harvard.edu/abs/2014Natur.515...85Z}
}

@ARTICLE{Simionescu2012,
       author = {{Simionescu}, A. and {Werner}, N. and {Urban}, O. and {Allen}, S.~W. and {Fabian}, A.~C. and {Sanders}, J.~S. and {Mantz}, A. and {Nulsen}, P.~E.~J. and {Takei}, Y.},
        title = "{Large-scale Motions in the Perseus Galaxy Cluster}",
      journal = {\apj},
         year = 2012,
        month = oct,
       volume = {757},
       number = {2},
          eid = {182},
        pages = {182},
          doi = {10.1088/0004-637X/757/2/182},
archivePrefix = {arXiv},
       eprint = {1208.2990},
 primaryClass = {astro-ph.CO},
       adsurl = {https://ui.adsabs.harvard.edu/abs/2012ApJ...757..182S}
}

@ARTICLE{Zuhone2011,
       author = {{ZuHone}, J.~A. and {Markevitch}, M. and {Lee}, D.},
        title = "{Sloshing of the Magnetized Cool Gas in the Cores of Galaxy Clusters}",
      journal = {\apj},
         year = 2011,
        month = dec,
       volume = {743},
       number = {1},
          eid = {16},
        pages = {16},
          doi = {10.1088/0004-637X/743/1/16},
archivePrefix = {arXiv},
       eprint = {1108.4427},
 primaryClass = {astro-ph.CO},
       adsurl = {https://ui.adsabs.harvard.edu/abs/2011ApJ...743...16Z}
}

@ARTICLE{Churazov2001,
       author = {{Churazov}, E. and {Br{\"u}ggen}, M. and {Kaiser}, C.~R. and {B{\"o}hringer}, H. and {Forman}, W.},
        title = "{Evolution of Buoyant Bubbles in M87}",
      journal = {\apj},
         year = 2001,
        month = jun,
       volume = {554},
       number = {1},
        pages = {261-273},
          doi = {10.1086/321357},
archivePrefix = {arXiv},
       eprint = {astro-ph/0008215},
 primaryClass = {astro-ph},
       adsurl = {https://ui.adsabs.harvard.edu/abs/2001ApJ...554..261C}
}

@ARTICLE{Fabian2000,
       author = {{Fabian}, A.~C. and {Sanders}, J.~S. and {Ettori}, S. and {Taylor}, G.~B. and {Allen}, S.~W. and {Crawford}, C.~S. and {Iwasawa}, K. and {Johnstone}, R.~M. and {Ogle}, P.~M.},
        title = "{Chandra imaging of the complex X-ray core of the Perseus cluster}",
      journal = {\mnras},
         year = 2000,
        month = nov,
       volume = {318},
       number = {4},
        pages = {L65-L68},
          doi = {10.1046/j.1365-8711.2000.03904.x},
archivePrefix = {arXiv},
       eprint = {astro-ph/0007456},
 primaryClass = {astro-ph},
       adsurl = {https://ui.adsabs.harvard.edu/abs/2000MNRAS.318L..65F}
}

@ARTICLE{Hogan2017,
       author = {{Hogan}, M.~T. and {McNamara}, B.~R. and {Pulido}, F. and {Nulsen}, P.~E.~J. and {Russell}, H.~R. and {Vantyghem}, A.~N. and {Edge}, A.~C. and {Main}, R.~A.},
        title = "{Mass Distribution in Galaxy Cluster Cores}",
      journal = {\apj},
         year = 2017,
        month = mar,
       volume = {837},
       number = {1},
          eid = {51},
        pages = {51},
          doi = {10.3847/1538-4357/aa5f56},
archivePrefix = {arXiv},
       eprint = {1610.04617},
 primaryClass = {astro-ph.GA},
       adsurl = {https://ui.adsabs.harvard.edu/abs/2017ApJ...837...51H}
}

@ARTICLE{ZuHone2016,
       author = {{ZuHone}, J.~A. and {Miller}, E.~D. and {Simionescu}, A. and {Bautz}, M.~W.},
        title = "{Simulating Astro-H Observations of Sloshing Gas Motions in the Cores of Galaxy Clusters}",
      journal = {\apj},
         year = 2016,
        month = apr,
       volume = {821},
       number = {1},
          eid = {6},
        pages = {6},
          doi = {10.3847/0004-637X/821/1/6},
archivePrefix = {arXiv},
       eprint = {1508.04426},
 primaryClass = {astro-ph.HE},
       adsurl = {https://ui.adsabs.harvard.edu/abs/2016ApJ...821....6Z}
}

@ARTICLE{Rhea2025,
       author = {{Rhea}, Carter Lee and {Hlavacek-Larrondo}, Julie and {Gendron-Marsolais}, Marie-Lou and {Vigneron}, Benjamin and {Donahue}, Megan and {Thilloy}, Auriane and {Rousseau-Nepton}, Laurie and {Mezcua}, Mar and {Werner}, Norbert and {Barrera-Ballesteros}, Jorge and {Choi}, Hyunseop and {Edge}, Alastair and {Fabian}, Andrew and {Voit}, G. Mark},
        title = "{Mapping the Filamentary Nebula of NGC 1275 with Multiwavelength SITELLE Observations}",
      journal = {\aj},
         year = 2025,
        month = apr,
       volume = {169},
       number = {4},
          eid = {203},
        pages = {203},
          doi = {10.3847/1538-3881/adb732},
archivePrefix = {arXiv},
       eprint = {2502.05406},
 primaryClass = {astro-ph.GA},
       adsurl = {https://ui.adsabs.harvard.edu/abs/2025AJ....169..203R}
}

@ARTICLE{Gendron2018,
       author = {{Gendron-Marsolais}, M. and {Hlavacek-Larrondo}, J. and {Martin}, T.~B. and {Drissen}, L. and {McDonald}, M. and {Fabian}, A.~C. and {Edge}, A.~C. and {Hamer}, S.~L. and {McNamara}, B. and {Morrison}, G.},
        title = "{Revealing the velocity structure of the filamentary nebula in NGC 1275 in its entirety}",
      journal = {\mnras},
         year = 2018,
        month = sep,
       volume = {479},
       number = {1},
        pages = {L28-L33},
          doi = {10.1093/mnrasl/sly084},
archivePrefix = {arXiv},
       eprint = {1802.00031},
 primaryClass = {astro-ph.GA},
       adsurl = {https://ui.adsabs.harvard.edu/abs/2018MNRAS.479L..28G}
}

@ARTICLE{Fabian2008,
       author = {{Fabian}, A.~C. and {Johnstone}, R.~M. and {Sanders}, J.~S. and {Conselice}, C.~J. and {Crawford}, C.~S. and {Gallagher}, III, J.~S. and {Zweibel}, E.},
        title = "{Magnetic support of the optical emission line filaments in NGC 1275}",
      journal = {\nat},
         year = 2008,
        month = aug,
       volume = {454},
       number = {7207},
        pages = {968-970},
          doi = {10.1038/nature07169},
archivePrefix = {arXiv},
       eprint = {0808.2712},
 primaryClass = {astro-ph},
       adsurl = {https://ui.adsabs.harvard.edu/abs/2008Natur.454..968F}
}

@ARTICLE{Tang2017,
       author = {{Tang}, Xiaping and {Churazov}, Eugene},
        title = "{Sound wave generation by a spherically symmetric outburst and AGN feedback in galaxy clusters}",
      journal = {\mnras},
         year = 2017,
        month = jul,
       volume = {468},
       number = {3},
        pages = {3516-3532},
          doi = {10.1093/mnras/stx590},
archivePrefix = {arXiv},
       eprint = {1701.05231},
 primaryClass = {astro-ph.GA},
       adsurl = {https://ui.adsabs.harvard.edu/abs/2017MNRAS.468.3516T}
}

@ARTICLE{Kirkpatrick2011,
       author = {{Kirkpatrick}, C.~C. and {McNamara}, B.~R. and {Cavagnolo}, K.~W.},
        title = "{Anisotropic Metal-enriched Outflows Driven by Active Galactic Nuclei in Clusters of Galaxies}",
      journal = {\apjl},
         year = 2011,
        month = apr,
       volume = {731},
       number = {2},
          eid = {L23},
        pages = {L23},
          doi = {10.1088/2041-8205/731/2/L23},
archivePrefix = {arXiv},
       eprint = {1103.0793},
 primaryClass = {astro-ph.GA},
       adsurl = {https://ui.adsabs.harvard.edu/abs/2011ApJ...731L..23K}
}

@ARTICLE{Simionescu_2026,
       author = {{Simionescu}, A. and {Kilbourne}, C. and {Russell}, H.~R. and {Ito}, D. and {Charbonneau}, M. and {Eckert}, D. and {Loewenstein}, M. and {Martin}, J. and {McCall}, H. and {McNamara}, B.~R. and {Nakazawa}, K. and {Ogorzalek}, A. and {T{\"u}mer}, A. and {Zhuravleva}, I. and {Dizdar}, N. and {Ezoe}, Y. and {Fujimoto}, R. and {Gu}, L. and {Hodges-Kluck}, E. and {Ichinohe}, Y. and {Kitamoto}, S. and {Leutenegger}, M.~A. and {Mernier}, F. and {Miller}, E.~D. and {Mitsuishi}, I. and {Sato}, K. and {Szymkowiak}, A.},
        title = "{Dynamics of AGN feedback in the X-ray bright East and Southwest arms of M87, mapped by XRISM}",
      journal = {arXiv e-prints},
         year = 2026,
        month = jan,
          eid = {arXiv:2601.16901},
        pages = {arXiv:2601.16901},
          doi = {10.48550/arXiv.2601.16901},
archivePrefix = {arXiv},
       eprint = {2601.16901},
 primaryClass = {astro-ph.GA},
       adsurl = {https://ui.adsabs.harvard.edu/abs/2026arXiv260116901S}
}

@article{Congyao2022,
    author = {Zhang, Congyao and Zhuravleva, Irina and Gendron-Marsolais, Marie-Lou and Churazov, Eugene and Schekochihin, Alexander A and Forman, William R},
    title = {Bubble-driven gas uplift in galaxy clusters and its velocity features},
    journal = {Monthly Notices of the Royal Astronomical Society},
    volume = {517},
    number = {1},
    pages = {616-631},
    year = {2022},
    month = {11},
    issn = {0035-8711},
    doi = {10.1093/mnras/stac2282},
    url = {https://doi.org/10.1093/mnras/stac2282},
    eprint = {https://academic.oup.com/mnras/article-pdf/517/1/616/46383813/stac2282.pdf},
}
\bibliographystyle{aasjournal}



\end{document}